\documentclass[manuscript,screen]{acmart}
\usepackage{amsmath,amsfonts,etoolbox}
\newtheorem{assumption}{Assumption}[section]
\newcommand{\eps}{\varepsilon}
\newcommand{\inv}{^{-1}}
\newcommand{\A}{\mathcal A}
\newcommand{\D}{\mathcal D}
\newcommand{\emptyword}{\epsilon}
\newcommand{\ellone}{\ell_1}
\newcommand{\elltwo}{\ell_2}
\acmJournal{TALG}
\setcopyright{none}
\copyrightyear{}
\acmDOI{}

\makeatletter
\patchcmd{\maketitle}{Manuscript submitted to ACM\\}{}{}%
  {\PackageError{manuscript}{Cannot remove the default submission-status line}{Check the ACM title block.}}%
\AtBeginDocument{%
  \apptocmd{\ps@standardpagestyle}{%
    \fancyfoot[RO,LE]{\footnotesize Manuscript for ACM}}{}%
    {\PackageError{manuscript}{Cannot set the draft footer}{Check the ACM class page styles.}}%
  \apptocmd{\ps@firstpagestyle}{%
    \fancyfoot[RE,LO]{\footnotesize Manuscript for ACM}}{}%
    {\PackageError{manuscript}{Cannot set the first-page draft footer}{Check the ACM class page styles.}}%
  \pagestyle{standardpagestyle}%
}
\makeatother
\title[Faster Treasure Hunt via Strongly Universal Exploration]{Faster Treasure Hunt via Strongly Universal Exploration with Polylogarithmic Overhead}
\titlenote{This manuscript extends the author's ISAAC 2007 paper, \emph{Faster Treasure Hunt and Better Strongly Universal Exploration Sequences}, pp.~549--560, DOI: 10.1007/978-3-540-77120-3\_48, and the 2012 preprint arXiv:1204.5442. The extensions include summable insertion periods and fixed-depth iterated-logarithm bounds, a complete integer-period window proof, direct instruction access with logarithmic-workspace recomputation, and a characterization of linear expansion.}
\author{Qin Xin}
\affiliation{\department{Faculty of Science and Technology}\institution{University of the Faroe Islands}\city{T\'orshavn}\country{Faroe Islands}}
\email{qinx@setur.fo}

\begin{abstract}
We study deterministic exploration and delayed-activation treasure hunt in anonymous undirected port-labeled graphs with incoming-port information. From polynomial-time constructible ordinary universal exploration sequences of length \(O(n^c)\) for a fixed regular degree, we construct a single infinite strongly universal exploration sequence with cover time \(O(n^c\log^2n)\). Summable dyadic insertion schedules yield the sharper bound \(O_\eps(n^c\log n(\log\log n)^{1+\eps})\) for every fixed \(\eps>0\), together with refinements at every fixed iterated-logarithm depth. The proof combines reversible excursions, neutral guards and an arbitrary-window argument that treats both orientations simultaneously. For this recurrence, uniform linear expansion is equivalent to convergence of the normalized period sum. Exact counting identities also permit direct instruction access. With a logspace ordinary generator, table-free enumeration of decreasing scale paths computes instruction \(t\) using \(O(\log(t+2))\) working bits and time polynomial in \(t\). The earlier polynomial-slack bounds follow as corollaries. A local cubic simulation extends the treasure-hunt application to arbitrary-degree graphs, with the \(2m\)-vertex replacement size included explicitly in the bound.
\end{abstract}
\ccsdesc[500]{Theory of computation~Graph algorithms analysis}
\ccsdesc[500]{Theory of computation~Distributed algorithms}
\ccsdesc[300]{Theory of computation~Design and analysis of algorithms}
\keywords{Strongly universal exploration sequences, deterministic treasure hunt, anonymous graphs, mobile agents, backtracking, graph exploration}
\begin{document}
\maketitle
\raggedbottom

\section{Introduction}
Universal exploration sequences encode deterministic walks that cover every graph in a specified class, for every local port numbering and every starting directed edge. Their strongly universal counterparts require the same guarantee in every contiguous window of a prescribed length. This distinction matters when a mobile agent has been running for an arbitrarily long time before a target becomes active: a covering prefix alone does not bound the delay after an arbitrary later activation.

The treasure hunt problem formalizes this requirement. An agent explores an unknown connected graph, while a stationary treasure is activated at an adversarially chosen time and vertex. The agent knows neither the graph size nor the treasure's activation time. The objective is to bound the meeting delay from the later activation in terms of the graph alone. A strongly universal exploration sequence provides such a bound when the agent learns its incoming port after each traversal.

Ta-Shma and Zwick introduced the delayed treasure hunt formulation and strongly universal sequences \cite{tz07,tz}. Their construction converts ordinary exploration sequences of length \(O(n^c)\) into strongly universal exploration sequences with cover time \(O(n^{2c})\). The author's conference paper \cite{xin07}, with the later presentation \cite{xin}, improved the exponent to \(c(1+1/\lambda)\) for each sufficiently large fixed integer \(\lambda\). The dependence on a fixed parameter is essential: one cannot infer a single near-linear-overhead infinite sequence merely by allowing that parameter to vary with the unknown graph size.

We replace polynomial slack by summable insertion schedules. The following two schedules give the principal bounds; each defines a single infinite sequence independently of the unknown graph size and activation time.
\begin{center}
\fbox{\begin{minipage}{\dimexpr\columnwidth-2\fboxsep-2\fboxrule\relax}
\begin{theorem}[Headline bounds]\label{thm:headline}
Assume polynomial-time ordinary UESs of length \(O(n^c)\) for a fixed regular degree, with \(c\geq1\).
\begin{enumerate}
\item The periods
\[
q_j^{\rm sq}=2^{j+4+\lceil2\log_2(j+1)\rceil}
\]
give one prefix-polynomially constructible infinite SUES with cover time \(O(n^c\log^2n)\).
\item Fix a rational \(\eps>0\), and put
\[
\begin{gathered}
h_j=(j+2)\lceil\log_2(j+2)\rceil^{1+\eps},\\
K_\eps=\max\{4,\lceil\log_2(8(1+1/\eps))\rceil\},\\
q_j^{\rm it}=2^{j+K_\eps+\lceil\log_2h_j\rceil}.
\end{gathered}
\]
These periods give one such SUES with cover time
\begin{equation}\label{eq:main}
O_\eps\!\left(n^c\log n(\log\log n)^{1+\eps}\right).
\end{equation}
\end{enumerate}
\end{theorem}
\end{minipage}}
\end{center}
Both statements follow from Theorem~\ref{thm:main} and the budget calculations in Section~\ref{sec:schedules}. For any real \(\eps>0\), a smaller fixed positive rational gives the asserted upper bound. Here \(c\) is inherited from the ordinary UES family, whereas \(\eps\) controls the logarithmic overhead.
To make all asymptotic expressions valid at small sizes, regularize each logarithm by \(1\). Write
\(\ell_1(x)=\max\{1,\log_2x\}\) and
\(\ell_{\nu+1}(x)=\max\{1,\log_2\ell_\nu(x)\}\). For \(r\geq1\), define
\[
\Phi_{r,\eps}(x)=
\left(\prod_{\nu=1}^{r-1}\ell_\nu(x)\right)\ell_r(x)^{1+\eps},
\]
where an empty product equals one. Theorem~\ref{thm:iterated-depth} gives the more general bound
\(O_{r,\eps}(n^c\Phi_{r,\eps}(n))\) at every fixed depth \(r\).

\subsection{Contributions and relation to the conference version}
The main contribution is the general summable-period conversion in Theorem~\ref{thm:main}. Its proof supplies two estimates: every finite recursive word expands its skeleton by at most a factor of five, and every window of length \(12q_j+3\) contains a complete covering word for scale \(2^j\). Both estimates apply to one fixed infinite sequence, and the window argument treats forward and inverse recursive copies simultaneously.

Table~\ref{tab:bounds} compares the conversion costs in the same fixed-degree model. Here \(L=n^c\), up to constant-factor normalization of the underlying UES family. The constants may depend on the displayed fixed parameter.
\begin{table}[t]
\centering
\caption{Cover-time bounds relative to an \(O(n^c)\)-length ordinary UES family.}\label{tab:bounds}
\begin{tabular}{ll}
\toprule
Construction & Cover time\\
\midrule
Ta-Shma--Zwick \cite{tz07,tz} & \(O(L^2)\)\\
Earlier conference result \cite{xin07,xin} & \(O_\lambda(L^{1+1/\lambda})\)\\
Summable square-log periods & \(O(L(\log L)^2)\)\\
Summable log-power periods & \(O_\eps(L(\log L)^{1+\eps})\)\\
Summable iterated-log periods & \(O_\eps(L\log L(\log\log L)^{1+\eps})\)\\
Any fixed depth \(r\) & \(O_{r,\eps}(L\Phi_{r,\eps}(L))\)\\
\bottomrule
\end{tabular}
\end{table}

The conversion admits two implementations with different time and space costs. Exact excursion counts yield direct instruction access with at most one skeleton query and, with a logspace base generator, \(O(\log^2(t+2))\) working bits at absolute position \(t\) (Theorem~\ref{thm:direct-access}). Table-free enumeration of decreasing scale paths reduces the workspace to \(O(\log(t+2))\) bits while retaining polynomial time in \(t\) (Theorem~\ref{thm:table-free} and Corollary~\ref{cor:low-space}). A complementary structural result, Theorem~\ref{thm:linear-criterion}, characterizes uniform linear expansion by convergence of the normalized period sum. Its harmonic-boundary consequence applies specifically to this recurrence.

The fixed-degree conversion preserves the exponent \(c\) of the ordinary UES family. Its treasure-hunt application extends to arbitrary-degree graphs through a local cubic simulation that charges one physical slot per virtual step and includes the \(2m\)-vertex replacement size in the bound. Both guarantees assume incoming-port information. The earlier \(O(n^{3c/2})\) and \(O_\lambda(n^{c(1+1/\lambda)})\) bounds follow as corollaries with exact integer periods. The existence of an \(O(n^c)\) cover-time conversion remains unresolved here.

All constructions and proofs of the new results are given in full. Appendix~\ref{app:legacy} analyzes the earlier unguarded words: an exact integer ruler interpretation gives the improved \(23M\) length bound for the original \(3/2\)-scale word, and a finite-tail argument proves linear length for every fixed polynomial-slack parameter. Appendix~\ref{app:series} gives a closed form for the historical auxiliary series, and Appendix~\ref{app:protocol} presents the complete search procedure. The ordinary UES family is used through the interface specified in Section~\ref{sec:model}.

\subsection{Related work}
Random walks and universal traversal sequences provide the classical background \cite{aleliunas,coppersmith}. Exploration sequences with backtracking were studied by Kouck\'y \cite{koucky}. Reingold's logspace construction supplies the explicit polynomial-length UES family used as a black box here \cite{reingold05,reingold08}. Earlier deterministic rendezvous algorithms, under several activation and information assumptions, include \cite{dessmark03,dessmark06,kowalski04,kowalski06}; asynchronous rendezvous is considered in \cite{demarco}. The present paper concerns the stationary-target application and does not assert a new label-dependent bound for two moving agents.

The 2026 journal study of self-stabilizing graph exploration \cite{sudo} discusses the earlier strongly universal exploration bounds, while its own algorithms permit node whiteboards and use additional information. Here the graph is anonymous and has no writable node memory. A bound from a model with whiteboards is therefore not directly comparable to the sequence conversion studied in this paper.

Space and marker tradeoffs for undirected graph exploration are studied by Disser, Hackfeld and Klimm \cite{disser19}. Their bounds use graph size and marker or agent resources. The present generation bound is in absolute instruction position, with no pebbles or node memory. Pelc \cite{pelc23} surveys the distinct label, activation and movement assumptions in deterministic rendezvous.

\section{Model and preliminaries}\label{sec:model}
\subsection{Graphs, instructions, and activation}
For the sequence theorems, fix \(d\geq3\). A graph is a nonempty, finite, connected, undirected \(d\)-regular port-labeled multigraph. Loops and parallel edges are allowed, as required by the cubic simulation. Every edge, including a loop, has two incident ports; local port numbers lie in \(\{0,\ldots,d-1\}\). The port numbers at the two ends of an edge need not agree.

A state is a directed edge ending at the current vertex. Its incident ports are part of the state, so parallel edges and the two orientations of a loop remain distinguishable as directed edges. If its incoming port is \(a\), instruction \(b\in\mathbb Z_d\) traverses the edge at port \(a+b\bmod d\). Thus an instruction word \(W=w_1\cdots w_t\) and a starting directed edge \(e_0=(v_{-1},v_0)\) determine states \(e_i=(v_{i-1},v_i)\), \(1\leq i\leq t\). Coverage counts the current vertex \(v_0\) and the vertices \(v_1,\ldots,v_t\); it does not count an unvisited tail \(v_{-1}\). An initial incoming port may be chosen fictitiously, since the guarantee holds for every starting directed edge.

Time is measured in discrete slots. Activations occur at slot boundaries, and a meeting includes the current vertex before the next move. Each edge traversal or stay consumes one slot; local computation is accounted for separately.

The agent knows the degree of its current vertex and learns the entry port after an edge traversal. It may keep internal state and may traverse an edge or stay in place in each synchronous slot. It has no vertex identifiers, markers, or writable node memory. The regular-graph sequence itself uses only edge traversals. The general-graph simulation may use stay slots. Time bounds count all slots.

For treasure hunt, the agent and the stationary treasure have arbitrary activation times. A meeting requires both to be active at the same vertex. The delay is measured from the later activation. The agent is not told the activation time of the treasure or the graph size.

\begin{definition}
A word is a UES for a graph class if its walk covers every graph in the class from every starting directed edge and every port labeling. An infinite word is a SUES with cover time \(p(n)\) if every contiguous factor of length \(p(n)\) is a UES for graphs of size at most \(n\).
\end{definition}
An infinite word is \emph{prefix-polynomially constructible} if its first \(T\) output symbols can be generated deterministically in time polynomial in the numerical output length \(T\). Period computation is governed separately by Assumption~\ref{ass:period-listing}; an oracle for the periods is not part of the constructibility definition. Traversal slots and local computation are analyzed as separate resources.

\subsection{Reversal and neutral excursions}
For a word \(W=w_1\cdots w_t\), let
\[
W\inv=(-w_t)\cdots(-w_1),\qquad
\D(W)=W\,0\,W\inv\,0 .
\]
All arithmetic on instructions is modulo \(d\).
\begin{lemma}\label{lem:reversal}
The words \(00\) and \(\D(W)\) restore their starting directed-edge state. Also
\(\D(W)\inv=0W0W\inv\). If \(W\) is a UES, then \(0W\inv\) is a UES under the coverage convention above.
\end{lemma}
\begin{proof}
Let \(F_W\) denote the state permutation induced by \(W\). Let \(J\) reverse a directed edge, and let \(R_b\) rotate an outgoing port by \(b\) at its tail. The state map of one instruction is \(R_bJ\), a permutation, and its inverse is \(JR_{-b}=J(R_{-b}J)J\). Multiplying these identities gives
\(F_{W\inv}=JF_W^{-1}J\).
The state map of \(\D(W)\) is therefore
\(JF_{W\inv}JF_W=F_W^{-1}F_W\), the identity. Also \(J^2\) is the identity, proving the assertion for \(00\). Reversing and negating the letters of \(\D(W)\) gives the displayed inverse word.

To prove coverage, take any starting directed edge \(e\) for \(0W\inv\). Because \(F_W\) is a permutation, there is a directed edge \(f\) such that executing \(W\) from \(f\) ends at \(e\). If this run visits \(v_0,\ldots,v_t\), the run of \(0W\inv\) from \(e\) starts at \(v_t\), traverses the same path backwards, and additionally reaches \(v_{-1}\). It therefore visits every \(v_i\), \(0\leq i\leq t\), at or after the start of the guarded inverse walk.
\end{proof}

\subsection{A prefix-compatible skeleton}
An ordinary UES family need not be prefix compatible. We first combine its words into a single infinite skeleton while preserving the polynomial length exponent.

Assume that ordinary UESs \(E_n\), universal for all graphs in the chosen fixed-degree class with at most \(n\) vertices, have length at most \(An^c\) and are deterministically constructible in polynomial time, for a fixed \(c\geq1\). Reingold's construction supplies such a fixed exponent \cite{reingold08}; its rotation-map framework permits the loops and parallel edges required below. Constants may depend on that fixed degree and chosen base family; displayed subscripts record additional fixed parameters.

If a base theorem is formulated for exactly \(n\) vertices, concatenate its outputs for sizes \(1,\ldots,N\) to obtain an at-most-\(N\) family. Every valid graph has its own covering component, entered from an arbitrary state. Polynomial length and time, and logarithmic streaming workspace, are preserved. This optional normalization may increase an unspecified polynomial exponent; throughout, \(c\) denotes the exponent of the chosen at-most-size family. The conversion adds no further power of \(n\).
\begin{lemma}\label{lem:skeleton}
There is a prefix-polynomially constructible infinite word \(U_\infty=u_1u_2\cdots\) such that, for every \(n\), some dyadic \(k\leq Cn^c\) has \(U_k=u_1\cdots u_k\) universal for graphs of size at most \(n\). Every \(U_k\) is a prefix of every longer \(U_M\).
\end{lemma}
\begin{proof}
For each \(h\geq0\), append zeros to \(E_{2^h}\), if necessary, until its length is at least \(2^h\). Its padded length is
\(\max\{|E_{2^h}|,2^h\}\leq(A+1)2^{hc}\), since \(c\geq1\). Concatenate these padded components for \(h=0,1,\ldots\).
For \(h=\lceil\log_2n\rceil\), the prefix through the complete component \(E_{2^h}\) has length
\[
\sum_{t=0}^{h}\max\{|E_{2^t}|,2^t\}
\leq(A+1)\sum_{t=0}^{h}2^{tc}
=O(2^{hc})=O(n^c).
\]
The component \(E_{2^h}\) covers every graph of size at most \(n\), regardless of the state reached through the preceding components. Hence this entire prefix is universal for those graphs.
Round this prefix length up to a power of two \(k\). Its extra suffix cannot destroy coverage, and \(k\) increases by less than a factor of two.

To generate a prefix of \(T\geq1\) symbols, the component for size \(2^h\) can be needed only if the preceding components have total length less than \(T\). Their padding gives total length at least \(2^h-1\), so \(2^h\leq T\). Thus all required component generators, including a possible final partial component, have inputs of size at most \(T\), and there are \(O(\log(T+1))\) components. Their construction and padding are polynomial in \(T\). The empty prefix is immediate. This normalization never needs to compute the exponent \(c\).
\end{proof}
We use these skeleton prefixes for every dyadic length \(M\). The graph size \(n\), the skeleton length \(M\), and the recursive target length \(k\) are distinct parameters (Table~\ref{tab:notation}). In the application, Lemma~\ref{lem:skeleton} chooses \(k\leq Cn^c\); \(M\) is a larger containing dyadic size and is not a graph-size parameter.

\begin{table}[t]
\caption{Parameters used in the conversion.}\label{tab:notation}
\centering
\begin{tabular}{@{}lp{0.71\columnwidth}@{}}
\toprule
Symbol&Meaning\\
\midrule
\(n\)&Number of graph vertices\\
\(M\)&Dyadic outer skeleton length of \(S_M\)\\
\(s_M\)&Expanded instruction length \(|S_M|\)\\
\(k=2^j\)&Dyadic target skeleton length\\
\(Q=q_j\)&Insertion period for target \(k\)\\
\(c,\eps\)&Base UES exponent; fixed logarithmic slack\\
\(t\)&Absolute instruction position\\
\bottomrule
\end{tabular}
\end{table}

\section{A general summable-period theorem}\label{sec:period-framework}
Choose a strictly increasing sequence of powers of two \(q_0,q_1,\ldots\) such that
\begin{equation}
q_j\geq16\cdot2^j,\qquad
\alpha:=\sum_{j\geq0}\frac{2^j}{q_j}\leq\frac18. \label{eq:periods}
\end{equation}
Because the periods are powers of two, each period divides every larger one. Define
\begin{equation}
r(i)=\max\bigl(\{\,2^j:q_j\mid i\,\}\cup\{0\}\bigr). \label{eq:selector}
\end{equation}
Here \(q_j\mid i\) means exact integer divisibility, not the inequality \(q_j\leq i\). Since \(q_j\geq16\cdot2^j\), every nonzero \(r(i)\) is at most \(i/16\).

Define recursively
\[
S_1=u_1,\qquad
D_i=
\begin{cases}
\emptyword,&r(i)=0,\\
S_{r(i)}\,0\,S_{r(i)}\inv\,0,&r(i)>0,
\end{cases}
\]
and, for dyadic \(M\geq2\),
\begin{equation}
S_M=u_1\,(00D_1u_2)\,(00D_2u_3)\cdots(00D_{M-1}u_M). \label{eq:construction}
\end{equation}
All referenced words are smaller than \(M\), so this definition is well founded. Neither \(r(i)\) nor the skeleton symbols depend on the terminal size \(M\); hence \(S_k\) is a prefix of \(S_M\) whenever \(k\leq M\) are dyadic. Let
\[
S_\infty=\lim_{a\to\infty}S_{2^a}.
\]

\begin{assumption}[Period listing]\label{ass:period-listing}
Given an integer \(B\geq1\), the complete indexed list of periods \(q_j\leq B\) can be computed in time polynomial in \(B\).
\end{assumption}
This is a separate algorithmic hypothesis, not part of the definition of prefix-polynomial constructibility. Every concrete schedule in Section~\ref{sec:schedules} satisfies it.

\begin{theorem}[Summable-period conversion]\label{thm:main}
For the construction \eqref{eq:periods}--\eqref{eq:construction}, the following hold.
\begin{enumerate}
\item Every dyadic word has length \(\lvert S_M\rvert\leq5M\).
\item For every \(j\geq0\), writing \(k=2^j\), every factor of \(S_\infty\) of length \(12q_j+3\) contains \(S_k\) or \(0S_k\inv\) as a contiguous factor.
\item If \(U_k\) is a UES for graphs of size at most \(n\), then \(S_\infty\) has cover time at most \(12q_j+3\) on those graphs.
\end{enumerate}
Finite prefixes are deterministically constructible in time polynomial in their length if the skeleton prefixes are and Assumption~\ref{ass:period-listing} holds.
\end{theorem}

The proof combines a summable length bound with endpoint control for arbitrary windows. The guard pair \(00\) is neutral on the directed-edge state and costs two symbols per skeleton gap. For dyadic \(R>k\), the construction gives \(S_R=S_k00V\), hence \(S_R^{-1}=V^{-1}00S_k^{-1}\). Therefore \(\D(S_R)^{-1}=0S_R0S_R^{-1}\) ends with the guarded target \(0S_k^{-1}\). When \(R=k\), the central reversal zero supplies the guard instead. Lemma~\ref{lem:guard} establishes the endpoint properties in both orientations.

\section{Linear expansion and boundary control}\label{sec:structural}
Write \(s_M=\lvert S_M\rvert\). We first bound the expansion at every recursive scale. This estimate will control both the lengths of endpoint targets and the total contribution of smaller insertions within one period.

\begin{lemma}[Linear expansion]\label{lem:length}
For every dyadic \(M\), \(s_M\leq5M\).
\end{lemma}
\begin{proof}
Induct on the dyadic size \(M\). The base case is \(s_1=1\leq5\).
For \(M\geq2\), the construction has \(M\) skeleton symbols and \(M-1\) skeleton gaps, with two guard zeros per gap. Excursion separators are counted inside \(2s_{r(i)}+2\). There are at most \(M/q_h\) positions whose selected value is \(r(i)=2^h\), since every such position is divisible by \(q_h\). Overcounting positions divisible by several periods only increases the bound. Using the induction hypothesis,
\begin{align*}
s_M
&=M+2(M-1)+\sum_{\substack{1\leq i<M\\r(i)>0}}(2s_{r(i)}+2)\\
&\leq3M+M\sum_{h:q_h<M}\frac{10\cdot2^h+2}{q_h}\\
&\leq3M+12M\sum_{h\geq0}\frac{2^h}{q_h}\\
&=(3+12\alpha)M\leq\tfrac92 M<5M.
\end{align*}
Every active child satisfies \(r(i)\leq i/16<M\), so each use of the induction hypothesis is valid.
\end{proof}

Fix a target scale \(k=2^j\), put \(Q=q_j\), and define the two target words
\[
\A_k=\{S_k,\ 0S_k\inv\}.
\]
Call a factor \emph{avoiding} if it contains neither target word. A top-level insertion \(\D(S_R)\), or its inverse, is \emph{large} at this scale if \(R\geq k\).
We next locate targets at the ends of these blocks and bound the gaps between them.

\begin{lemma}[Guards at both ends of large blocks]\label{lem:guard}
Each large block, in either orientation, contains a target word within its first \(B=s_k+2\) symbols and within its last \(B\) symbols. Furthermore, \(B\leq Q\).
\end{lemma}
\begin{proof}
For dyadic \(R>k\), prefix compatibility and \eqref{eq:construction} give \(S_R=S_k\,00\,V\) for some word \(V\). Thus \(S_R\inv=V\inv\,00\,S_k\inv\).
Consequently \(\D(S_R)\) begins with \(S_k\) and ends with \(0S_k\inv0\). For equality \(R=k\), explicitly,
\(\D(S_k)=S_k\,0S_k^{-1}0\) begins with \(S_k\) and ends with \(0S_k^{-1}0\), while
\(\D(S_k)^{-1}=0S_k\,0S_k^{-1}\) begins with \(0S_k\) and ends with \(0S_k^{-1}\).
For \(R>k\), since
\[
\D(S_R)\inv=0\,S_R\,0\,S_R\inv,
\]
the inverse block begins with \(0S_k\) and ends with \(0S_k\inv\).
The stated occurrences fit within \(B\) symbols of both ends. Finally,
since \(k\geq1\) and \(Q\geq16k\) by \eqref{eq:periods},
\[
B=s_k+2\leq5k+2\leq7k\leq\frac7{16}Q<\frac Q2.
\]
In particular, the bound \(B\leq Q\) follows solely from
\eqref{eq:periods} and Lemma~\ref{lem:length}.
\end{proof}

\begin{lemma}[Short gaps]\label{lem:gaps}
In the top-level decomposition of \(S_M\), or of \(S_M\inv\), every gap outside large blocks has length at most \(5Q\).
\end{lemma}
\begin{proof}
If \(M\leq Q\), the entire word has length at most \(5M\leq5Q\), by Lemma~\ref{lem:length}. Hence suppose \(M>Q\).
An insertion is large exactly when its index is divisible by \(Q\). Indeed, divisibility by a higher period implies divisibility by \(Q\), and divisibility by \(Q\) activates level \(j\).
The delimiters are successive multiples of \(Q\), so their index interval contains at most \(Q\) skeleton symbols. There are at most \(Q\) associated skeleton transitions, each carrying one guard pair: at most \(2Q\) guard zeros. Boundary gaps satisfy the same bounds. All remaining excursions have scales \(h<j\); their two separator zeros are counted separately inside \(2s_{2^h}+2\).

Since every \(q_h\), \(h<j\), divides \(Q\), any \(Q\) consecutive index positions contain exactly \(Q/q_h\) multiples of \(q_h\). An interval of at most \(Q\) positions contains no more. Overcounting as in Lemma~\ref{lem:length}, the gap length is at most
\[
3Q+\sum_{h<j}\frac{Q}{q_h}(2s_{2^h}+2)
\leq(3+12\alpha)Q<5Q.
\]
Inverting a word reverses these gaps and preserves their lengths.
\end{proof}

\begin{lemma}[Short avoiding prefixes and suffixes]\label{lem:ends}
For every dyadic \(M\), each avoiding prefix or suffix of \(S_M\), or of \(S_M\inv\), has length at most \(6Q\).
\end{lemma}
\begin{proof}
If \(M\leq Q\), any prefix or suffix, including the whole word, has length
at most \(s_M\leq5M\leq5Q<6Q\), by Lemma~\ref{lem:length}.
Otherwise \(M\geq2Q\), because both are dyadic.
At indices \(Q\) and \(M-Q\), the selected value in \eqref{eq:selector} is exactly \(k\): their 2-adic valuation equals that of \(Q\), and all higher periods are multiples of \(2Q\).
There is therefore a full \(\D(S_k)\) at each of these indices, with the same block serving both roles when \(M=2Q\).

The first and last gaps each have length at most \(5Q\), by Lemma~\ref{lem:gaps}. By Lemma~\ref{lem:guard}, a target occurs within the first \(B\) symbols of the first large block and within the last \(B\) symbols of the last large block. An avoiding prefix or suffix must stop before it contains the corresponding occurrence, so its length is at most \(5Q+B\leq6Q\). The same argument applies to the inverted blocks in \(S_M\inv\).
\end{proof}

\section{The arbitrary-window guarantee}\label{sec:window}
The boundary estimates control an avoiding factor that crosses top-level block boundaries. A factor contained in a single large block is handled by induction on its smaller recursive halves. We state the boundary case first to keep these two arguments independent.

\begin{lemma}[Meeting ordered blocks]\label{lem:block-meeting}
Fix \(k,Q\) as above. An avoiding factor not contained in one top-level large block meets at most two large blocks. If it meets two, they are consecutive, and its intersections are a suffix of the first and a prefix of the second. Its length is at most \(2B+5Q\leq7Q\).
\end{lemma}
\begin{proof}
The top-level large blocks occupy disjoint ordered intervals. A contiguous factor meeting three of them contains the middle interval entirely, including a target from Lemma~\ref{lem:guard}. This contradicts avoidance. The argument uses only the endpoint targets.

If two blocks are met, any intervening block would likewise be fully contained; hence the two are consecutive. The factor cannot contain either whole block, so its intersections are a suffix and a prefix. Each has length less than \(B\): an intersection of length at least \(B\) would contain the corresponding endpoint target. The intervening gap has length at most \(5Q\) by Lemma~\ref{lem:gaps}. If one or no block is met, the same estimate holds with the absent pieces omitted. A factor meeting one block cannot extend beyond both its ends, since it would then contain that block entirely.
\end{proof}

\begin{lemma}[Window lemma]\label{lem:windows}
Every avoiding factor of \(S_M\) or \(S_M\inv\) has length at most \(12Q+2\).
\end{lemma}
\begin{proof}
Induct simultaneously on the dyadic size \(M\), for both orientations.
For \(M=1\), each orientation consists of one symbol, so every avoiding
factor has length at most \(1\leq12Q+2\).
Now let \(M\geq2\), and assume the claim for every smaller dyadic size in
both orientations. If there are no large blocks, Lemma~\ref{lem:gaps}
bounds the entire word, and hence every factor, by \(5Q\).

If an avoiding factor \(T\) is not contained in a single large block, Lemma~\ref{lem:block-meeting} gives \(|T|\leq7Q\).

Otherwise \(T\) lies inside \(\D(S_R)=S_R0S_R\inv0\), or inside its inverse \(0S_R0S_R\inv\), where \(R<M\).
If \(T\) is contained in one of the two halves \(S_R,S_R\inv\), the induction hypothesis applies.
In every remaining case, \(T\) contains at least one separator zero. Its nonempty intersections with the halves are avoiding prefixes or suffixes, each of length at most \(6Q\) by Lemma~\ref{lem:ends}. There are at most two separator zeros. The same estimate covers a factor consisting only of a separator, so \(\lvert T\rvert\leq12Q+2\) in all cases.
\end{proof}

\begin{proof}[Proof of Theorem~\ref{thm:main}]
The length claim is Lemma~\ref{lem:length}. Any finite factor of the prefix limit \(S_\infty\) is contained in some \(S_M\), so Lemma~\ref{lem:windows} gives the claimed target in every length-\(12Q+3\) window.

By Lemma~\ref{lem:reversal}, every \(00\) and every \(D_i\) in \eqref{eq:construction} restores its starting directed-edge state. Consequently, at its designated skeleton steps, the execution of \(S_k\) follows exactly the walk prescribed by \(U_k\). If \(U_k\) is a UES, then \(S_k\) covers every required vertex as well. The same lemma makes \(0S_k\inv\) a UES.
A word containing either target as a contiguous factor therefore covers the graph from every starting directed edge. This establishes the SUES property.

To generate a prefix of \(T\geq1\) symbols, choose a dyadic \(M\) with \(T\leq M<2T\), and generate \(S_M\) recursively with cached smaller words. It contains at least \(M\) symbols and at most \(5M\). The sum of the lengths of all cached dyadic words is \(O(M)\); a straightforward implementation, including period listing, divisibility tests, and skeleton generation, is polynomial in \(T\). This proves the stated explicitness.
\end{proof}

\section{Explicit schedules and the resulting bounds}\label{sec:schedules}
The conversion theorem reduces the cover-time analysis to a choice of periods. We now give explicit schedules, verify their summability budgets and establish the required integer computability.

\subsection{A square-logarithm schedule}\label{subsec:square}
Choose
\[
q_j=2^{\,j+4+\lceil2\log_2(j+1)\rceil}.
\]
Then
\[
\begin{gathered}
16\cdot2^j(j+1)^2\leq q_j<32\cdot2^j(j+1)^2,\\
\alpha\leq\frac1{16}\sum_{j\geq0}\frac1{(j+1)^2}
\leq\frac1{16}\left(1+\int_1^\infty x^{-2}\,dx\right)=\frac18.
\end{gathered}
\]
The first periods are \(16,128,1024,2048,8192,\ldots\).
Since the relevant dyadic \(k=2^j\) can be chosen to be \(O(n^c)\), Theorem~\ref{thm:main} gives \(O(n^c\log^2 n)\).

\subsection{Any fixed power above one on the logarithm}\label{subsec:logpower}
For fixed \(\eps>0\), let
\[
\begin{gathered}
K_\eps=\max\!\left\{4,\left\lceil\log_2\bigl(8(1+1/\eps)\bigr)\right\rceil\right\},\\
q_j=2^{\,j+K_\eps+\lceil(1+\eps)\log_2(j+1)\rceil}.
\end{gathered}
\]
The integral test gives
\[
\alpha\leq 2^{-K_\eps}\sum_{j\geq0}(j+1)^{-1-\eps}
\leq2^{-K_\eps}(1+1/\eps)\leq\tfrac18.
\]
The periods satisfy \eqref{eq:periods}, so the cover time is
\[
O_\eps\!\left(n^c(\log n)^{1+\eps}\right).
\]
For algorithmic definiteness one may choose any fixed positive rational \(\eps\); for an arbitrary requested real exponent, a smaller positive rational gives a bound at least as strong.

\subsection{A two-level iterated-logarithm schedule}\label{subsec:iterlog}
More generally, take any nondecreasing \(h_j\geq1\) with
\(\sum_jh_j^{-1}<\infty\), and choose a fixed integer \(K\geq4\) satisfying
\(2^{-K}\sum_jh_j^{-1}\leq1/8\). Put
\[
q_j=2^{\,j+K+\lceil\log_2h_j\rceil}.
\]
These are strictly increasing dyadic periods, satisfy \eqref{eq:periods}, and are \(O(2^jh_j)\). For the explicitness claim, require that the complete list of periods at most \(B\) can be computed in time polynomial in \(B\). The concrete schedules below satisfy this condition.
Taking
\[
h_j=(j+2)\bigl(\lceil\log_2(j+2)\rceil\bigr)^{1+\eps}
\]
makes the reciprocal sum converge and yields \eqref{eq:main}. More explicitly, grouping the integers \(t=j+2\) by \(m=\lceil\log_2t\rceil\) gives
\[
\sum_{j\geq0}\frac1{h_j}
\leq\sum_{m\geq1}\frac1{m^{1+\eps}}
\leq1+1/\eps.
\]
Thus the same \(K_\eps\) as in Section~\ref{subsec:logpower} is sufficient. For fixed rational \(\eps=p/q>0\), the rounding test \(2^v\geq h_j\) is exactly the integer comparison
\(2^{vq}\geq(j+2)^q\lceil\log_2(j+2)\rceil^{p+q}\).
This computes the integer periods without floating-point ambiguity in polynomial time. Smaller rational parameters cover every requested real \(\eps>0\), as in Section~\ref{subsec:logpower}.

\subsection{Any fixed iterated-logarithm depth}\label{subsec:any-depth}
The same construction admits an arbitrary fixed number of iterated logarithms. The depth is fixed when the infinite sequence is defined.

\begin{theorem}[Fixed-depth iterated-logarithm conversion]\label{thm:iterated-depth}
Assume the ordinary UES family of Lemma~\ref{lem:skeleton}. For every fixed integer \(r\geq1\) and every fixed \(\eps>0\), there is a prefix-polynomially constructible infinite SUES with cover time
\[
O_{r,\eps}\!\left(n^c\Phi_{r,\eps}(n)\right).
\]
For \(r=1,2,3\), respectively, the overheads are
\[
\begin{array}{c|l}
r&\text{overhead}\\ \hline
1&\ell_1(n)^{1+\eps}\\
2&\ell_1(n)\ell_2(n)^{1+\eps}\\
3&\ell_1(n)\ell_2(n)\ell_3(n)^{1+\eps}.
\end{array}
\]
\end{theorem}
\begin{proof}
For positive integers \(t\), define the rounded logarithms
\[
\begin{gathered}
\Lambda_1(t)=\max\{1,\lceil\log_2t\rceil\},\\
\Lambda_{\nu+1}(t)=\max\{1,\lceil\log_2\Lambda_\nu(t)\rceil\}.
\end{gathered}
\]
For \(r=1\), use \(h_j=(j+1)^{1+\eps}\). For \(r\geq2\), put \(t=j+2\) and set
\[
h_j=t\left(\prod_{\nu=1}^{r-2}\Lambda_\nu(t)\right)
             \Lambda_{r-1}(t)^{1+\eps}.
\]
Each \(h_j\) is nondecreasing and at least one.

We first bound the reciprocal sum. Let \(B_s\), \(s\geq2\), denote the sum of the reciprocals of the displayed depth-\(s\) weights over \(t\geq2\). Group \(t\) by \(m=\Lambda_1(t)\). For every \(m\geq1\), the group is exactly
\(t=2^{m-1}+1,\ldots,2^m\), and
\[
\sum_{\substack{t\geq2\\\Lambda_1(t)=m}}\frac1t
=\sum_{t=2^{m-1}+1}^{2^m}\frac1t
<\frac{2^m-2^{m-1}}{2^{m-1}}=1.
\]
Also \(\Lambda_{\nu+1}(t)=\Lambda_\nu(m)\) on that group. Hence
\[
B_2\leq\sum_{m\geq1}m^{-1-\eps}\leq1+1/\eps,\qquad
B_s\leq1+B_{s-1}\quad(s\geq3).
\]
The extra one is the term \(m=1\); all its rounded logarithms equal one. Therefore, for every \(r\geq1\),
\[
\sum_{j\geq0}h_j^{-1}
\leq C_{r,\eps}:=\max\{1,r-1\}+1/\eps.
\]
For \(r=1\), the same bound follows directly from the integral test.

Choose
\[
K_{r,\eps}=\max\{4,\lceil\log_2(8C_{r,\eps})\rceil\},\qquad
q_j=2^{j+K_{r,\eps}+\lceil\log_2h_j\rceil}.
\]
These periods are strictly increasing powers of two, are at least \(16\cdot2^j\), and satisfy
\[
\sum_{j\geq0}\frac{2^j}{q_j}
\leq2^{-K_{r,\eps}}C_{r,\eps}\leq\frac18.
\]
For \(k=2^j\), rounded and regularized logarithms differ by only depth-dependent constant factors. Thus
\(h_j=O_{r,\eps}(\Phi_{r,\eps}(k))\) and
\(q_j=O_{r,\eps}(k\Phi_{r,\eps}(k))\).
Lemma~\ref{lem:skeleton} gives \(k=O(n^c)\); for \(1\leq\nu\leq r\), \(\ell_\nu(k)=O_{c,C,\nu}(\ell_\nu(n))\), where \(C\) is the fixed normalization constant from that lemma. Theorem~\ref{thm:main} gives the claimed cover time.

For explicitness, first take \(\eps=p/q>0\) rational. Write a weight as \(h_j=A_jB_j^{1+p/q}\), with \(A_j,B_j\) positive integers: for \(r=1\), use \(A_j=1,B_j=j+1\); for \(r\geq2\), use the integer product and final rounded logarithm above. The condition \(2^v\geq h_j\) is exactly
\[
2^{vq}\geq A_j^qB_j^{p+q}.
\]
All rounded logarithms and this comparison are computable with integer arithmetic. At fixed depth and fixed rational parameter, \(h_j\) is at most polynomial in \(j+2\). To list periods below \(B\geq1\), it suffices to examine \(j\leq\lfloor\log_2B\rfloor\), because \(q_j\geq16\cdot2^j\). Every such candidate period has \(O_{r,\eps}(\log(B+2))\) bits, so the list is computable in polynomial time. For a general real \(\eps>0\), choose a fixed positive rational parameter smaller than \(\eps\); its bound is at least as strong. The sequence and constants may depend on \(r,\eps\), but neither parameter depends on the unknown graph size or activation time.
\end{proof}

The summability requirement explains why this particular proof does not immediately give \(O(n^c\log n)\): periods \(q_j=\Theta(2^jj)\) make the charging sum divergent. This is a limitation of the present sufficient condition, not a lower bound for SUESs.

\section{Recovery of the earlier polynomial-slack bounds}\label{sec:earlier}
The period theorem also gives a direct proof of the asymptotic results in the conference paper and its preprint.
\begin{corollary}\label{cor:lambda}
For every fixed integer \(\lambda\geq1\), there is a prefix-polynomially constructible infinite SUES with cover time \(O_\lambda(n^{c(1+1/\lambda)})\). In particular, \(\lambda=2\) gives \(O(n^{3c/2})\).
\end{corollary}
\begin{proof}
Set \(K_\lambda=\lceil\log_2(16\lambda)\rceil\), and choose the exact dyadic periods
\[
q_j=2^{\,K_\lambda+\lceil(1+1/\lambda)j\rceil}.
\]
They are strictly increasing and at least \(16\cdot2^j\). To check the budget, let
\(f(x)=1-2^{-x}\). Since \(f''(x)=-(\ln2)^2 2^{-x}<0\), concavity on \([0,1]\) gives the chord bound
\[
f(x)\geq(1-x)f(0)+xf(1)=\frac x2
\qquad(0\leq x\leq1).
\]
Taking \(x=1/\lambda\) yields
\(1-2^{-1/\lambda}\geq1/(2\lambda)\). Hence
\[
\sum_{j\geq0}\frac{2^j}{q_j}
\leq\frac{2^{-K_\lambda}}{1-2^{-1/\lambda}}
\leq\frac{2\lambda}{16\lambda}=\frac18.
\]
For \(k=2^j\), \(q_j<2^{K_\lambda+1}k^{1+1/\lambda}\). Apply Theorem~\ref{thm:main} and Lemma~\ref{lem:skeleton}. The periods use only integer arithmetic because \(\lambda\) is an integer. Each fixed \(\lambda\) defines one infinite sequence valid at all graph sizes.
\end{proof}
The polylogarithmic schedules are asymptotically stronger than every fixed-\(\lambda\) bound. They are established directly by a single summable schedule; they do not follow from a size-dependent choice of \(\lambda\) in Corollary~\ref{cor:lambda}.

\section{Treasure hunt and arbitrary-degree graphs}\label{sec:applications}
The arbitrary-window guarantee applies directly at the later activation of the agent and treasure. We first state this consequence for the fixed-degree model, then give a local simulation for arbitrary degrees.

\begin{corollary}[Delayed treasure hunt]\label{cor:treasure}
For any cover time established by Theorem~\ref{thm:main}, executing the corresponding SUES finds an activated stationary treasure within that many slots after the later activation, on the fixed-degree graph class. For every fixed \(r,\eps\) as in Theorem~\ref{thm:iterated-depth}, the bound is \(O_{r,\eps}(n^c\Phi_{r,\eps}(n))\); in particular \(r=2\) gives \(O_\eps(n^c\ellone(n)\elltwo(n)^{1+\eps})\).
\end{corollary}
\begin{proof}
If the treasure activates later, consider the instruction window beginning at its activation. If the agent activates later, consider the window beginning with the agent's first instruction. In both cases the current directed edge is arbitrary and the next window is universal. The treasure vertex is visited while both entities are active. No knowledge of \(n\) or of either activation difference is used.
\end{proof}

We next make the degree reduction explicit. Let \(c_3\) denote the exponent of an ordinary UES family for connected cubic port-labeled multigraphs. We keep this symbol separate from any exponent for a different base graph class.

\begin{proposition}[Local cubic simulation]\label{prop:simulation}
Let \(G\) be a connected undirected port-labeled graph with \(m\geq1\) edges and arbitrary positive vertex degrees. For every fixed \(r\geq1,\eps>0\), and without knowing \(m\) or \(n\), an agent with incoming-port information can execute a cubic SUES virtually, one virtual step per physical slot. Thus deterministic treasure hunt is possible within
\[
O_{r,\eps}\!\left(m^{c_3}\Phi_{r,\eps}(m)\right)
\]
slots after the later activation. For simple \(n\)-vertex graphs this implies
\(O_{r,\eps}(n^{2c_3}\Phi_{r,\eps}(n))\).
\end{proposition}
\begin{proof}
Replace a vertex \(v\) of degree \(\delta_v\) by virtual vertices \((v,p)\), \(0\leq p<\delta_v\). Each virtual vertex has ports \(0,1,2\). Port \(0\) connects \((v,p)\) to \((v,p-1\bmod\delta_v)\), at port \(2\) of the latter. Port \(2\) is the reverse cycle direction. If original port \(p\) of \(v\) is paired with port \(q\) of \(w\), connect port \(1\) of \((v,p)\) to port \(1\) of \((w,q)\). Degrees one and two produce loops or parallel cycle edges, which are allowed.

This graph has \(\sum_v\delta_v=2m\) virtual vertices, is cubic, and is connected: each local cycle connects all ports at its vertex, and the cross edges reproduce every original adjacency.

The agent stores its current virtual index \(p\) and its virtual incoming port \(a\). At a cubic instruction \(b\), let \(z=a+b\bmod3\).
If \(z=0\), it stays at \(v\), replaces \(p\) by \(p-1\bmod\delta_v\), and sets \(a=2\).
If \(z=2\), it stays at \(v\), replaces \(p\) by \(p+1\bmod\delta_v\), and sets \(a=0\).
If \(z=1\), it traverses original port \(p\), learns entry port \(q\) at \(w\), and sets \(p=q\), \(a=1\).
Each case uses one slot and exactly simulates the relevant virtual edge. Initially choose \(p=0,a=1\), a valid fictitious incoming edge. All updates use only local degrees, the learned incoming port, and internal state.

The virtual sequence is followed continuously from activation. Every covering window for \(2m\) virtual vertices visits a virtual copy of every original vertex. Corollary~\ref{cor:treasure} therefore applies, including activation during a virtual cycle step, whose physical stay still consumes a slot. Theorem~\ref{thm:iterated-depth} supplies the cubic cover time at every fixed depth. Since \(2m=O(n^2)\) for simple graphs and fixed-depth regularized logarithms change by at most constant factors under a fixed polynomial rescaling, the final assertion follows. A connected graph with \(m=0\) has a single vertex and is immediate.
\end{proof}

This proposition is not an identification of \(c_3\) with a numerical exponent for arbitrary-degree UESs. The regularization size is part of the bound. For graphs with parallel edges, \(m\) need not be \(O(n^2)\), so the edge-parameter bound is the applicable statement.

\section{Exact lengths and direct access}\label{sec:random-access}
The prefix construction proves explicitness by materializing finite words. A different implementation can locate a single instruction without storing the expanded word. This distinction is relevant when an agent has been active for a long time: the graph-traversal guarantee is independent of that age, while an implementation should avoid retaining a record of all preceding instructions.

We derive exact counting identities that determine the expansion geometry and allow navigation through the recursive word. Lemmas~\ref{lem:exact-count} and~\ref{lem:boundary-position} require only strictly increasing dyadic periods with \(q_h\geq16\cdot2^h\); their proofs do not use summability. The resource bounds in this section additionally assume the period budget of Theorem~\ref{thm:main}. A skeleton query returns a symbol \(u_i\) of the normalized infinite base word. We separate the cost of that query from the additional cost of the conversion.

\subsection{Counting selected excursions}
For a positive integer \(L\), define
\begin{equation}\label{eq:exact-count}
N_h(L)=
\left\lfloor\frac{L-1}{q_h}\right\rfloor
-\left\lfloor\frac{L-1}{q_{h+1}}\right\rfloor .
\end{equation}
The upper limit \(L-1\) matters: a word ending at its \(L\)th skeleton symbol does not include the excursion following that symbol. A period exceeding this limit contributes a zero quotient, so only finitely many terms can be nonzero.

\begin{lemma}[Exact excursion counts]\label{lem:exact-count}
Precisely \(N_h(L)\) indices in \(1,\ldots,L-1\) select \(r(i)=2^h\). Moreover,
\[
\sum_{h\geq0}N_h(L)=
\left\lfloor\frac{L-1}{q_0}\right\rfloor .
\]
\end{lemma}
\begin{proof}
The condition \(r(i)\geq2^h\) is equivalent to \(q_h\mid i\). Since the higher periods are nested multiples of \(q_{h+1}\), the condition \(r(i)>2^h\) is equivalent to \(q_{h+1}\mid i\). Their difference counts the selected level exactly. The nonnegative differences telescope, and the final quotient is zero at sufficiently high levels. In contrast to the earlier upper estimates, this count does not charge an index to every period dividing it.
\end{proof}

Let \(F(L)\) be the one-based position of the \(L\)th designated outer skeleton symbol in the expanded infinite word. Designation refers to the outer skeleton \(U_\infty\), not to additional copies of a letter inside excursions. Prefix compatibility makes the position independent of the dyadic terminal size used to contain it.

\begin{lemma}[Skeleton boundary positions]\label{lem:boundary-position}
For every positive integer \(L\),
\begin{equation}\label{eq:boundary-position}
\begin{split}
F(L)={}&3L-2\\
 &+2\sum_{h\geq0}N_h(L)(s_{2^h}+1).
\end{split}
\end{equation}
For dyadic \(M\), \(F(M)=s_M\), and
\[
F(L+1)-F(L)=
\begin{cases}
3,&r(L)=0,\\
3+2(s_{r(L)}+1),&r(L)>0.
\end{cases}
\]
In particular, \(F\) is strictly increasing.
\end{lemma}
\begin{proof}
The prefix through the \(L\)th designated symbol has \(L\) skeleton letters, \(L-1\) guard pairs, and exactly the excursions counted by Lemma~\ref{lem:exact-count}. A level-\(h\) excursion has length \(2s_{2^h}+2\). Adding the three contributions gives \eqref{eq:boundary-position}. At dyadic terminal size \(M\), the prefix is exactly \(S_M\).

Passing to the next skeleton boundary appends two guard zeros, the selected excursion if nonempty, and one skeleton letter. This gives the difference formula. At \(L=1\) the sum vanishes and \(F(1)=1\), so the smallest case is included.
\end{proof}

This definition avoids assigning a recursive word to a nondyadic size. The quantity \(F(L)\) identifies a boundary inside a word already defined by the construction; it does not require extending the family \(S_M\) by an unstated rounding rule.

\subsection{Length tables without expanded words}
Fix \(M=2^a\) and put \(T_b=s_{2^b}\). The exact recurrence is
\begin{equation}\label{eq:length-table}
\begin{split}
T_0={}&1,\\
T_b={}&3\cdot2^b-2\\
 &+2\sum_{h:q_h<2^b}N_h(2^b)(T_h+1),
 \qquad b\geq1 .
\end{split}
\end{equation}
A nonzero summand has \(q_h<2^b\) and \(q_h\geq16\cdot2^h\). Therefore \(h\leq b-5\), and each required child length is available before its parent is evaluated.

\begin{proposition}[Length preprocessing]\label{prop:length-preprocessing}
Given the indexed periods below \(M\), the table \(T_0,\ldots,T_a\) can be computed with \(O((a+1)^2)\) arithmetic operations on \(O(a+1)\)-bit integers and \(O((a+1)^2)\) bits of storage.
\end{proposition}
\begin{proof}
Evaluate \eqref{eq:length-table} in increasing order of \(b\). Each of at most \(a\) stages has at most \(a\) relevant levels. Quotients by the dyadic periods are shifts. A next period beyond the bound contributes zero and can be represented by a sentinel rather than by an arbitrarily large integer.

The linear bound gives \(T_b\leq5\cdot2^b\). Counts, products and partial sums consequently have \(O(a+1)\) bits. There are \(O(a+1)\) lengths and periods to retain. The operation and storage bounds follow. These are arithmetic-operation counts on integers of the stated bit length; elementary integer algorithms give a polynomial bit bound as well.
\end{proof}

The tables depend only on the insertion periods. They describe the expansion geometry independently of the ordinary exploration generator and can be reused for any skeleton word over a fixed-degree instruction alphabet.

\subsection{Locating a symbol}
For \(1\leq p\leq s_M\), binary search with \eqref{eq:boundary-position} finds the largest \(L\leq M\) with \(F(L)\leq p\). Equality identifies \(u_L\). Otherwise put \(\delta=p-F(L)\). Offsets one and two are guard zeros.

For \(\delta>2\), the selected excursion is nonempty. Set \(o=\delta-2\), \(R=r(L)\), and \(v=s_R\). The offset is inside \(S_R0S_R^{-1}0\), with interpretation
\begin{equation}\label{eq:offset-map}
\begin{array}{c|c}
\text{offset}&\text{instruction}\\ \hline
1\leq o\leq v & S_R[o]\\
o=v+1 &0\\
v+2\leq o\leq2v+1 &-S_R[2v+2-o]\\
o=2v+2 &0 .
\end{array}
\end{equation}
The reflected index is always in \(1,\ldots,v\): its values at the two endpoints are \(v\) and \(1\), respectively.

\begin{figure}[t]
\centering
\begin{minipage}{\columnwidth}
\hrule\smallskip
\textbf{Direct instruction query}\\
Input: dyadic \(M\), position \(p\), length table.\\
Initialize the sign \(\sigma=1\).
\begin{enumerate}
\item Find the largest \(L\leq M\) with \(F(L)\leq p\).
\item If \(p=F(L)\), return \(\sigma u_L\bmod d\).
\item Set \(\delta=p-F(L)\). If \(\delta\leq2\), return zero.
\item Put \(o=\delta-2\), \(R=r(L)\), \(v=s_R\).
\item If \(o=v+1\) or \(o=2v+2\), return zero.
\item If \(o\leq v\), replace \((M,p)\) by \((R,o)\).
\item Otherwise replace
\[
(M,p,\sigma)\ \text{by}\ (R,2v+2-o,-\sigma).
\]
\item Repeat from step 1.
\end{enumerate}
\smallskip\hrule
\end{minipage}
\caption{One-symbol access without retaining an expanded word. In step 2, \(d\) is the fixed graph degree.}
\Description{An iterative instruction query uses binary search on skeleton boundaries, returns a guard zero directly, and otherwise descends into a smaller forward child or reflects an inverse-child index while reversing its sign. Only the terminal letter case queries the normalized skeleton.}
\label{fig:direct-query}
\end{figure}

Figure~\ref{fig:direct-query} is iterative. Its sign records containing inverse copies. After a child is selected, the parent is never used again, so no stack of parent expansions is needed.

\begin{theorem}[Direct-access conversion]\label{thm:direct-access}
Given the periods and length table for \(M=2^a\), one symbol of \(S_M\) can be found with at most one skeleton query, \(O((a+1)^3)\) additional arithmetic operations, and \(O((a+1)^2)\) additional working bits. For position \(t\) in the infinite word, \(a=O(\log(t+2))\) suffices.
\end{theorem}
\begin{proof}
Maintain the invariant that the requested original symbol is \(\sigma S_M[p]\). The monotonic boundary function uniquely identifies the containing skeleton interval. At equality it returns its skeleton letter. The next two positions are the guards.

A remaining position must lie in a nonempty excursion. Indeed, an empty excursion leaves only the two guards before the next designated letter, so \(\delta>2\) would contradict the maximal choice of \(L\). The four rows of \eqref{eq:offset-map} are disjoint and exhaust the excursion. The forward row preserves the sign; the inverse row reflects the position and changes the sign. Both preserve the invariant. A separator returns zero irrespective of the sign.

Every nonterminal iteration replaces \(M\) by \(R=r(L)\leq L/16<M/16\). There are only \(O(a+1)\) iterations. At \(M=1\), the unique valid position is the skeleton boundary, proving termination also at the smallest size. Only a terminal letter case calls the skeleton oracle, so there is at most one such query.

Evaluating \(F(L)\) takes \(O(a+1)\) arithmetic operations. Binary search uses \(O(a+1)\) evaluations per iteration; a scan of the same period list selects \(r(L)\). Thus the total extra cost is \(O((a+1)^3)\) operations. Each integer has \(O(a+1)\) bits. The tables use quadratic storage and the current search state uses linear further storage.

For an infinite-word position \(t\geq1\), choose the smallest dyadic \(M\geq t\). Since \(s_M\geq M\), the query is valid. Prefix compatibility makes the result independent of that terminal choice. The inequality \(t\leq M<2t\) gives the claimed logarithmic parameter bound, including \(t=1\).
\end{proof}

The theorem bounds the cost of navigating the recursive word independently of the ordinary exploration generator. The cost of the normalized skeleton query must be added separately. Thus the conversion itself avoids storing or generating an expanded word whose length is comparable to the absolute instruction index.

\subsection{Length recomputation in linear workspace}\label{subsec:table-free}
The quadratic-bit table is useful for fast queries, but it is not necessary for polynomial-time generation. Recomputing each required length permits a smaller workspace. The recomputation enumerates subsets of the \emph{scale indices}, rather than expanded instructions.

For every \(b\geq0\), define
\[
\begin{split}
g_b
&=3\cdot2^b-2+2\sum_{h\geq0}N_h(2^b)\\
&=3\cdot2^b-2+
 2\left\lfloor\frac{2^b-1}{q_0}\right\rfloor ,
\end{split}
\]
where the second equality is the telescoping identity in
Lemma~\ref{lem:exact-count}. For \(0\leq h<b\), put
\(c_{b,h}=2N_h(2^b)\). Separating the two contributions in
\(2N_h(2^b)(T_h+1)\) rewrites the exact recurrence as
\begin{equation}\label{eq:affine-length}
T_b=g_b+\sum_{h=0}^{b-1}c_{b,h}T_h.
\end{equation}
The sum is empty when \(b=0\), giving \(T_0=g_0=1\).
All levels, including \(h=0\), contribute their separator costs to \(g_b\).
The separate term \(c_{b,0}T_0\), when nonzero, accounts for the two copies
of \(S_1\) per level-zero excursion. Thus the level-zero contribution is
neither omitted nor counted twice. Nonzero coefficients have \(h\leq b-5\).

For a subset \(J\subseteq\{0,\ldots,b-1\}\), write its members in decreasing order \(h_1>\cdots>h_\ell\), and put \(h_0=b\). Set
\[
w_b(J)=g_{h_\ell}\prod_{\nu=1}^{\ell}c_{h_{\nu-1},h_\nu}
\quad(J\ne\varnothing),\qquad
w_b(\varnothing)=g_b.
\]
A product with a zero coefficient contributes zero.

\begin{lemma}[Subset-path length formula]\label{lem:subset-length}
For every \(b\geq0\),
\begin{equation}\label{eq:subset-length}
T_b=\sum_{J\subseteq\{0,\ldots,b-1\}}w_b(J).
\end{equation}
Given read-only access to the relevant dyadic periods, \(T_b\), for \(b\leq a\), can be evaluated using \(O(a+1)\) additional working bits and \(2^b(a+1)^{O(1)}\) bit operations apart from period-access costs.
\end{lemma}
\begin{proof}
Unroll \eqref{eq:affine-length}. Every nonconstant term selects a strictly smaller child index. A complete term therefore follows a decreasing path \(b=h_0>h_1>\cdots>h_\ell\), ending at the constant \(g_{h_\ell}\). Such a path is uniquely specified by its set of child indices. The empty path contributes \(g_b\). Conversely, grouping nonempty subsets by their largest member recovers the recurrence, so induction on \(b\) proves \eqref{eq:subset-length}, including \(b=0\).

Enumerate the subsets by a \(b\)-bit integer mask from zero through \(2^b-1\). For each mask scan the possible child indices in decreasing order. Retain only the current index and coefficient product, and finally multiply by the endpoint constant. A zero coefficient permits immediate rejection of that mask. Accumulate the resulting nonnegative terms. No list of paths, recursion stack, or array of earlier lengths is stored.

The expansion bound gives \(T_b\leq5\cdot2^b\). All terms in
\eqref{eq:subset-length} are nonnegative, so each complete term and each
partial accumulated sum is at most \(T_b\). For a nonzero
prefix product along \(b=h_0>h_1>\cdots>h_\nu\), put
\(P=\prod_{i=1}^{\nu}c_{h_{i-1},h_i}\).
Ending the path at this prefix gives the subset term
\(w_b(\{h_1,\ldots,h_\nu\})=P g_{h_\nu}\). Since every \(g_h\geq1\),
\[
P\leq P g_{h_\nu}
=w_b(\{h_1,\ldots,h_\nu\})\leq T_b.
\]
The initial empty product is one and is also at most \(T_b\).
Thus all retained arithmetic values have \(O(a+1)\) bits. The mask uses at
most \(a\) bits, and the index counters use \(O(\log(a+2))\).
Only a constant number of these integers is retained.

For length computations with \(b\leq a\), a period above \(2^a\) gives a zero quotient and can be replaced by a dyadic sentinel \(2^{a+1}\). All other periods have \(O(a+1)\) bits. Each of the \(2^b\) masks uses at most \(b\) coefficient computations and arithmetic on integers of this bit length. Elementary multiplication and division give the stated bit-operation bound.
\end{proof}

For example, at \(b=5\) under the square-log schedule, \(g_5=96\), \(c_{5,0}=2\), and all other outgoing coefficients vanish. The two nonzero subset terms give \(T_5=96+2g_0=98\), matching the first nonempty insertion in Section~\ref{sec:worked}.

\begin{theorem}[Table-free instruction access]\label{thm:table-free}
Suppose that, given \(a\) and \(0\leq h\leq a+1\), the capped exponent
\[
\min\{\log_2q_h,a+1\}
\]
is computable with \(O(a+1)\) working bits and \((a+1)^{O(1)}\) bit operations. Given \(M=2^a\) and a valid position \(p\) in \(S_M\), one instruction is computable using \(O(a+1)\) working bits, at most one skeleton query, and \(2^a(a+1)^{O(1)}\) additional bit operations.
\end{theorem}
\begin{proof}
Use the iterative query in Figure~\ref{fig:direct-query}, but evaluate every child length by Lemma~\ref{lem:subset-length} whenever it is needed. To evaluate \(F(L)\), scan its at most \(a+1\) count terms, recomputing each required \(T_h\) and adding its contribution. The period routine supplies exact divisors, with the sentinel convention of the preceding lemma. Hence all boundary values, separator decisions, and reflected indices remain unchanged.

The boundary routine retains only its argument, partial sum and current scale; its invocation of the length routine has constant call depth. The outer query retains only the current size, position, sign and binary-search state. A child descent discards the parent state, exactly as in Theorem~\ref{thm:direct-access}. Thus a constant number of routines, each with \(O(a+1)\) working bits, suffices; there is no stack of recursive word queries. The sign invariant and the at-most-one-skeleton-query property are unchanged.

One boundary evaluation uses at most \(a+1\) subset enumerations. Their combined time is bounded by \(2^a(a+1)^{O(1)}\). The outer procedure makes \(O((a+1)^2)\) boundary evaluations and at most \(O(a+1)\) child-length and ruler computations. Multiplying these bounds only changes the polynomial in \(a+1\). Period computation satisfies the same bound by hypothesis. This proves both the time and workspace claims.
\end{proof}

\subsection{Streaming generation and the time--space tradeoff}
Every concrete fixed-parameter schedule in this paper satisfies the capped-period hypothesis of Theorem~\ref{thm:table-free}. For the iterated-log schedules at fixed depth and rational \(\eps\), the integer factors defining \(h_j\) are polynomially bounded in \(j+2\). The exact rational-power comparisons in Section~\ref{subsec:any-depth} use \(O(\log(a+2))\)-bit integers for \(j\leq a+1\). The resulting period exponent is \(O(a+1)\) and can be capped before forming its power of two. The square-log and fixed-\(\lambda\) periods are simpler integer formulas. All use \(O(a+1)\) bits and polynomial time in \(a+1\); the weaker period-listing assumption alone would not imply this resource bound for an arbitrary schedule.

\begin{corollary}[Logarithmic-workspace streaming]\label{cor:low-space}
Assume that the ordinary UES family has a streaming generator using \(O(\log(n+2))\) working bits and time polynomial in \(n\). Every guarded SUES obtained from the square-log, rational-slack, fixed-depth or fixed-\(\lambda\) period formulas then admits instruction access at position \(t\) with \(O(\log(t+2))\) working bits and time polynomial in \(t\). Its first \(T\) instructions can also be emitted online with \(O(\log(T+2))\) working bits and time polynomial in \(T\), without knowing a final stopping length in advance.
\end{corollary}
\begin{proof}
The normalized skeleton has a logarithmic-workspace symbol routine. To query \(u_i\), stream the padded ordinary components of Lemma~\ref{lem:skeleton} until reaching the requested symbol. A needed component has size parameter at most \(i\), because its preceding padded components contain at least that parameter minus one symbols. If a base routine expects a unary input, its input can be simulated from the stored binary parameter and an input-head counter. The component parameter, remaining target offset, output counter and base-generator configuration use \(O(\log(i+2))\) bits. Logarithmically many component generators, each of polynomial time at a size at most \(i\), suffice.

For position \(t\geq1\), choose the least dyadic \(M\geq t\), so that \(t\leq M<2t\). Theorem~\ref{thm:table-free} supplies an \(O(\log(t+2))\)-bit query routine, whose only skeleton query has index at most \(M\). Its additional time is \(t(\log(t+2))^{O(1)}\), and the skeleton query takes polynomial time in \(t\).

To stream the word, maintain a counter \(t=1,2,\ldots\) and emit the value returned by this routine at each index. The counter needs only logarithmic further workspace. Summing a fixed polynomial per-query bound over the first \(T\) indices remains polynomial in \(T\). Prefix compatibility makes the emitted symbols independent of the successive containing sizes \(M\).
\end{proof}

The subset enumeration is polynomial in the \emph{numerical} instruction index \(t\), not in its bit length: \(2^a<2t\). Recomputing lengths therefore trades additional time for reduced storage. The table-based procedure still offers \(O(\log^3(t+2))\) additional arithmetic operations with \(O(\log^2(t+2))\) working bits; the table-free procedure offers logarithmic workspace with polynomial-index computation. Both use at most one ordinary-skeleton symbol query.

Reingold's streaming construction supplies a base family with the required resource properties. These bounds concern instruction generation. A physical simulator additionally retains its local port and controller state, and arbitrarily late activation can make \(t\) arbitrarily large even on a fixed graph.

\section{A characterization of linear expansion}\label{sec:linear-criterion}
The budget \(1/8\) in Theorem~\ref{thm:main} gives the explicit expansion factor five. We now remove that numerical budget and ask when the same recurrence has any uniform linear bound. Convergence of the normalized period sum gives an exact characterization.

\begin{theorem}[Linear-expansion criterion]\label{thm:linear-criterion}
For any strictly increasing dyadic periods satisfying \(q_h\geq16\cdot2^h\), construct the guarded words by \eqref{eq:construction}. Then
\[
\sup_{M\ {\rm dyadic}}\frac{s_M}{M}<\infty
\quad\Longleftrightarrow\quad
\sum_{h\geq0}\frac{2^h}{q_h}<\infty .
\]
This is a structural equivalence. Explicit generation additionally requires the relevant period-computability assumption.
\end{theorem}
\begin{proof}
For necessity, use the exact identities of Lemmas~\ref{lem:exact-count} and~\ref{lem:boundary-position}, which hold without a summability assumption. Assume \(s_M\leq CM\) at every dyadic size, and fix a finite \(H\). At sufficiently large dyadic \(M\), all periods through \(q_{H+1}\) divide \(M\) and are smaller than \(M\). Therefore
\[
\frac{N_h(M)}{M}
=\frac1{q_h}-\frac1{q_{h+1}}
\quad(0\leq h\leq H).
\]
Use the exact length formula, discard all higher nonnegative terms, and use \(s_{2^h}\geq2^h\). Letting \(M\) increase gives
\[
C\geq3+
2\sum_{h=0}^{H}2^h
 \left(\frac1{q_h}-\frac1{q_{h+1}}\right).
\]
The finite sum telescopes:
\begin{equation}\label{eq:criterion-telescope}
\begin{split}
&\sum_{h=0}^{H}2^h
 \left(\frac1{q_h}-\frac1{q_{h+1}}\right)\\
&\quad=\frac1{q_0}
+\frac12\sum_{h=1}^{H}\frac{2^h}{q_h}
-\frac{2^H}{q_{H+1}} .
\end{split}
\end{equation}
The boundary fraction satisfies \(0\leq2^H/q_{H+1}\leq1/32\), since \(q_{H+1}\geq16\cdot2^{H+1}\). Divergence of the normalized period sum would therefore make the right side of \eqref{eq:criterion-telescope} unbounded as \(H\to\infty\), contradicting the fixed bound \(C\).

For sufficiency, choose \(H\) with
\[
\beta=2\sum_{h\geq H}\frac{2^h}{q_h}<1.
\]
The finitely many words at levels below \(H\) have finite lengths. Put
\[
B=3+2\sum_{h<H}\frac{s_{2^h}+1}{q_h}+\beta
\]
and choose a finite constant \(C\) satisfying
\[
C\geq1,\qquad C\geq\frac{B}{1-\beta},\qquad
C\geq\max_{0\leq b<H}\frac{s_{2^b}}{2^b}.
\]
The last condition is omitted when \(H=0\).

Induct on the dyadic exponent of \(M\), with \(M=1\) and all exponents below \(H\) covered by \(C\). Count each selected level by all multiples of its period. For \(h<H\) use the fixed lengths; for \(h\geq H\) apply induction to the smaller active children. Hence
\[
\begin{split}
\frac{s_M}{M}
&\leq3+2\sum_{h<H}\frac{s_{2^h}+1}{q_h}\\
&\quad+
2C\sum_{h\geq H}\frac{2^h}{q_h}
+2\sum_{h\geq H}\frac1{q_h}\\
&\leq B+\beta C\leq C .
\end{split}
\]
Indeed, \(2\sum_{h\geq H}q_h^{-1}\leq\beta\), since \(2^h\geq1\), so the additive tail is included in \(B\). This completes the induction and proves uniform linearity.
\end{proof}

The sufficient direction tolerates finitely many expensive small scales. Their costs enter \(B\); convergence makes the remaining tail contractive. Given a certified upper bound on the normalized period sum, a common dyadic enlargement of all periods can make the budget at most \(1/8\), giving the explicit factor five. Convergence alone need not supply an effective bound from which to compute that enlargement.

\begin{corollary}[The harmonic boundary]\label{cor:harmonic-boundary}
If \(q_h=\Theta(2^h(h+1))\), the words do not have uniformly linear length. At all sufficiently large dyadic \(M\),
\[
s_M=\Omega(M\log\log M).
\]
\end{corollary}
\begin{proof}
Divergence of the harmonic series already gives the first assertion. For the quantitative estimate, write \(M=2^a\), \(H=\lfloor a/2\rfloor\), and choose \(C_q\) with \(q_h\leq C_q2^h(h+1)\). For sufficiently large \(a\), \(q_{H+1}<M\), so all counts through \(H\) have their exact normalized values.

The length lower bound and \eqref{eq:criterion-telescope} imply
\[
\begin{split}
\frac{s_M}{M}
&\geq-O(1)+\sum_{h=1}^{H}\frac{2^h}{q_h}\\
&\geq-O(1)+\frac1{C_q}\sum_{h=1}^{H}\frac1{h+1}.
\end{split}
\]
The harmonic sum is \(\Theta(\log H)=\Theta(\log a)=\Theta(\log\log M)\), giving the claimed lower bound.
\end{proof}

This is a barrier to linear expansion of the stated insertion recurrence. It is not a lower bound for arbitrary SUESs, other conversions, or a different analysis of nonlinearly expanding words. In particular, it does not rule out an \(O(n^c\log n)\) cover time by another method.

\section{Worked construction and boundary cases}\label{sec:worked}
The examples in this section use distinct formal skeleton letters. They illustrate the conversion independently of the base UES family. They are not assertions that the displayed short words alone explore a graph class.

\subsection{The first nonempty insertion}
For the square-log schedule, \(q_0=16\). No nonempty excursion occurs among indices \(1,\ldots,15\), and
\[
S_{16}=u_1\,00\,u_2\,00\cdots00\,u_{16},
\qquad s_{16}=46 .
\]
At index 16 the selector is \(r(16)=1\). The next gap therefore contains
\[
00\,S_1\,0\,S_1^{-1}\,0
=00\,u_1\,0\,(-u_1)\,0 .
\]
The gap occurs in \(S_{32}\), but not at the end of \(S_{16}\). There is only one nonempty excursion before index 32, so
\[
s_{32}=3\cdot32-2+2(s_1+1)=98 .
\]

\begin{table}[t]
\caption{Positions around the first selected excursion.}
\label{tab:first-excursion}
\centering
\begin{tabular}{cll}
\toprule
Position&Symbol&Role\\
\midrule
46&\(u_{16}\)&skeleton boundary\\
47,48&\(0,0\)&guards\\
49&\(u_1\)&forward child\\
50&\(0\)&reversal separator\\
51&\(-u_1\)&inverse child\\
52&\(0\)&return separator\\
53&\(u_{17}\)&next boundary\\
\bottomrule
\end{tabular}
\end{table}

The boundary formula gives \(F(17)=53\), agreeing with Table~\ref{tab:first-excursion}. A query at positions 47 or 48 terminates as a guard. Position 49 descends to \(S_1[1]\) with positive sign; position 51 reaches the same child with negative sign. Positions 50 and 52 use the separator cases.

The example also distinguishes a skeleton index from an expanded-word position. Index 16 selects the child, but its designated letter is at expanded position 46. A program that tests divisibility using the expanded position would be implementing a different, unproved sequence.

\begin{proposition}[An explicit square-log constant]\label{prop:square-constant}
For the concrete square-log schedule, \(s_M\leq(27/7)M\) at every dyadic size.
\end{proposition}
\begin{proof}
The normalized tail beyond level zero satisfies
\[
\beta:=2\sum_{h\geq1}\frac{2^h}{q_h}
\leq\frac18\sum_{h\geq1}(h+1)^{-2}\leq\frac18.
\]
The last inequality follows by comparison with the integral of \(x^{-2}\) from one to infinity. With \(H=1\), \(s_1=1\) and \(q_0=16\), the sufficiency proof of Theorem~\ref{thm:linear-criterion} gives
\[
B=3+\frac{2(s_1+1)}{16}+\beta\leq\frac{27}{8}.
\]
Consequently \(B/(1-\beta)\leq(27/8)/(1-1/8)=27/7\).
The choice \(C=27/7\) covers the initial size and closes the induction.
\end{proof}

\subsection{Rounded periods and fixed parameters}
Table~\ref{tab:period-examples} compares exact periods at small target scales. The depth-two and depth-three columns use \(\eps=1\), with the shifts given by their infinite-sum budget certificates.

\begin{table}[t]
\centering
\caption{Exact small-scale periods. A higher depth can have a larger finite constant.}
\label{tab:period-examples}
\centering
\begin{tabular}{rrrrr}
\toprule
Level&Target&Square-log&Depth two&Depth three\\
\midrule
0&1&16&32&64\\
1&2&128&512&512\\
2&4&1024&1024&1024\\
3&8&2048&8192&16384\\
4&16&8192&16384&65536\\
5&32&32768&32768&131072\\
6&64&65536&131072&262144\\
7&128&131072&524288&1048576\\
\bottomrule
\end{tabular}
\end{table}

A higher fixed depth need not improve the finite constants at every small scale. It can require a larger shift to preserve the entire budget, and its advantage concerns asymptotic growth after that shift and the depth have been fixed. The theorem is not a prescription for choosing the depth from an unknown graph size.

The dyadic rounding is structural. It ensures that every larger period is a multiple of every smaller one. Replacing a period by a nearby arbitrary integer could invalidate the large-block equivalence, the one-period gap count, and the ruler lookup. The analysis explicitly pays the constant factor incurred by rounding upward to a power of two.

\subsection{Activation beside an inverse boundary}
A window may start in an inverse excursion or immediately beside a skeleton symbol. It is not necessarily aligned with a forward copy. The target lemma must therefore work in both orientations and across separators.

For \(R>k\), the identity \(S_R=S_k00V\) makes \(0S_k^{-1}\) a suffix of the inverse, even when \(V\) contains many other scales. For \(R=k\), there is no intervening \(V\); the central reversal zero supplies the guard. Equality and strict inequality are separate cases in the endpoint proof for this reason.

At outer size \(M=2Q\), the first and last large blocks coincide. Both ends of that one block must be useful. At \(M\leq Q\), no such block need exist: the linear bound controls the whole word. Dyadicity ensures that these alternatives leave no missing intermediate-size case.

Coverage counts only vertices visited at or after a window begins. The target \(0S_k^{-1}\) therefore includes its leading zero. It is not justified by assuming that the tail of an incoming edge was visited while the treasure was already active. This convention is retained in the inversion proof and the physical simulator.

\subsection{Formal letters and actual exploration instructions}
After the construction, formal letters can be identified with actual values in \(\mathbb Z_d\). Such an identification may introduce additional target occurrences, but it cannot remove a literal occurrence already guaranteed by the word proof. Distinct formal letters therefore establish containment independently of the chosen instruction alphabet.

The ordinary UES theorem supplies graph coverage for every port labeling and starting directed edge. The examples show how reversible insertions and boundary indexing preserve that guarantee.

\section{Explicit generation}\label{sec:explicit}
For a requested prefix length \(T\geq1\), choose a dyadic \(M\) with \(T\leq M<2T\). Compute all \(S_{2^h}\), \(2^h\leq M\), in increasing order. At each skeleton position \(i\), append \(u_i\); if \(i<M\), append \(00\) and the excursion selected by the periods dividing \(i\). Every referenced word has size at most \(i/16\), so it has already been computed.

The sum of the cached expanded lengths is at most
\(5\sum_{h\leq\log_2M}2^h<10M\). A simple implementation may test divisibility against every listed period. There are \(O(\log M)\) possible levels below \(M\). Together with the period-listing and skeleton-generation assumptions, this is polynomial in \(T\). Inverse words can be copied by a reverse scan with negation. The construction requires no search over graphs.

The materializing method tests \(O(\log M)\) candidate periods per outer skeleton index. Copying an excursion can output many symbols, and the cache uses \(O(M)\) symbol storage. The instruction-access procedures in Section~\ref{sec:random-access} avoid this cache.

Theorem~\ref{thm:direct-access} gives a separate per-symbol procedure: at absolute position \(t\), at most one normalized-skeleton query and \(O(\log^3(t+2))\) additional arithmetic operations on \(O(\log(t+2))\)-bit integers, after period and length preprocessing. For the concrete schedules, that preprocessing has polynomial bit cost in \(\log(t+2)\). The table-free alternative in Theorem~\ref{thm:table-free} recomputes lengths. With a logspace streaming base generator it uses \(O(\log(t+2))\) working bits, at most one skeleton query, and time polynomial in \(t\). These are age-dependent computation costs; traversal rounds are still the separate graph-dependent measure.

\section{Discussion}
Theorem~\ref{thm:linear-criterion} identifies the precise threshold for uniform linear expansion of this recurrence. At the harmonic boundary \(q_j=\Theta(2^j(j+1))\), the exact count formula forces length \(\Omega(M\log\log M)\). Thus these words cannot satisfy a uniform \(O(M)\) bound. Allowing nonlinear expansion or changing the insertion grammar may lead to a different cover-time tradeoff. The existence of an \(O(n^c\log n)\) conversion remains open here.

The two instruction-access procedures offer a separate time--space tradeoff. Subset enumeration uses \(O(\log(t+2))\) working bits and time polynomial in the absolute instruction position \(t\). The memoized procedure reduces the additional arithmetic cost to \(O(\log^3(t+2))\) operations using an \(O(\log^2(t+2))\)-bit table. These generation costs can depend on an arbitrarily late activation, even though the traversal-delay bound depends only on the graph. Obtaining a comparable computation guarantee independent of the activation delay remains open here.

For two moving agents, covering every vertex in each agent's own window does not ensure simultaneous occupation of a vertex. One sufficient coordination mechanism would align an exploring interval of one agent with a stationary interval of the other lasting at least the relevant cover time. In a label-based rendezvous protocol, a label schedule must force such an overlap despite unknown activation offsets and graph size. A faster SUES supplies the exploration part; it does not supply that alignment or its label-dependent time analysis.

\section{Conclusion}\label{sec:conclusion}
For every fixed \(\eps>0\) and regular degree, we convert an \(O(n^c)\)-length ordinary UES family into a single infinite SUES with cover time \(O_\eps(n^c\ellone(n)\elltwo(n)^{1+\eps})\). Summable schedules give refinements at every fixed iterated-logarithm depth, and convergence of the normalized period sum exactly characterizes uniform linear expansion of this recurrence. With a logspace ordinary generator, exact excursion counts permit logarithmic-workspace instruction access in time polynomial in the absolute instruction position. The explicit cubic simulation extends delayed treasure hunt to arbitrary-degree graphs under the incoming-port model, with the \(2m\)-vertex replacement size included in the bound. A natural next question is whether nonlinear expansion or a different insertion scheme can achieve \(O(n^c\log n)\) cover time.

\begin{acks}
ChatGPT was used as a research and writing assistant during preparation of this extended version. The author proposed the summable-period construction and the fixed-depth refinements; ChatGPT assisted with refining these constructions and with suggesting clearer ways to present them, assisted with the derivation of the exact-count identities, the subset-path length formula, the logarithmic-workspace streaming procedure and the linear-expansion criterion, examined proofs and boundary cases, and improved the exposition and \LaTeX{} presentation. The research problem, core ideas, algorithmic designs and principal scientific contributions of the original conference paper were independently developed by the author. The author is responsible for the final manuscript and all scientific claims.
\end{acks}


\appendix

\section{Historical note: lengths in the earlier construction}\label{app:legacy}
This historical note is not used in the main conversion theorem or its window proof. It records the earlier linear-length assertions under an exact integer ruler interpretation and their improved constant. The unguarded legacy words below are separate from the guarded words in the main construction.

For a fixed integer \(\lambda\geq1\), put \(\theta=\lambda/(\lambda+1)\) and
\[
\rho(i)=2^{\lfloor\theta\nu_2(i)\rfloor},
\]
where \(\nu_2(i)\) is the exponent of the largest power of two dividing \(i\). Equivalently, \(\rho(i)\geq2^j\) exactly when \(2^{\lceil j/\theta\rceil}\mid i\). This interpretation specifies integer periods, is independent of the terminal size, and has \(\rho(M-i)=\rho(i)\) for dyadic \(M\) and \(1\leq i<M\).

Let \(R_1=u_1\), and use the earlier unguarded recurrence
\[
R_M=u_1\D(R_{\rho(1)})u_2\D(R_{\rho(2)})\cdots
u_{M-1}\D(R_{\rho(M-1)})u_M .
\]
Let \(a_h=|R_{2^h}|\) and \(A_h=a_h/2^h\).
\begin{lemma}\label{lem:legacy-recurrence}
For \(h\geq1\),
\[
\begin{gathered}
a_h=2^h+\sum_{p=0}^{h-1}2^{h-p-1}(2a_{\lfloor\theta p\rfloor}+2),\\
A_h\leq3+\sum_{p=0}^{h-1}2^{\lfloor\theta p\rfloor-p}A_{\lfloor\theta p\rfloor}.
\end{gathered}
\]
\end{lemma}
\begin{proof}
Exactly \(2^{h-p-1}\) indices \(i\in\{1,\ldots,2^h-1\}\) have \(\nu_2(i)=p\). Such an index selects \(R_{2^{\lfloor\theta p\rfloor}}\) and contributes two copies and two zeros. Divide by \(2^h\) and use \(\sum_{p\geq0}2^{-p}=2\). Every child exponent is less than \(h\).
\end{proof}
\begin{proposition}\label{prop:legacy-linear}
For every fixed \(\lambda\), there is a finite constant \(C_\lambda\) such that \(|R_M|\leq C_\lambda M\) at every dyadic \(M\).
\end{proposition}
\begin{proof}
Since \(p\) is an integer,
\(w_p=2^{\lfloor\theta p\rfloor-p}=2^{-\lceil p/(\lambda+1)\rceil}\).
Choose
\[
\begin{gathered}
L=\lceil\log_2(2(\lambda+2))\rceil,\quad P=(\lambda+1)L,\\
\delta=(\lambda+2)2^{-L}\leq\tfrac12 .
\end{gathered}
\]
The term at \(p=P\) is \(2^{-L}\). For each \(t\geq L+1\), exactly \(\lambda+1\) later indices have \(\lceil p/(\lambda+1)\rceil=t\). Hence the exact tail is
\[
\sum_{p\geq P}w_p
=2^{-L}+(\lambda+1)\sum_{t\geq L+1}2^{-t}
=\delta.
\]
Set \(H=\lfloor\theta(P-1)\rfloor=\lambda L-1\). The finitely many values \(A_0,\ldots,A_H\) are computable from Lemma~\ref{lem:legacy-recurrence}, without reference to later values. Put
\[
\begin{gathered}
B=3+\sum_{p=0}^{P-1}w_pA_{\lfloor\theta p\rfloor},\\
C_\lambda=\max\left\{A_0,\ldots,A_H,\frac{B}{1-\delta}\right\}.
\end{gathered}
\]
For \(h>H\), split the sum in the preceding lemma at \(P\). The initial part is at most \(B\), including its constant term; the remaining children are smaller than \(h\), so induction bounds the tail by \(C_\lambda\sum_{p\geq P}w_p\leq C_\lambda\delta\).
Thus \(A_h\leq B+C_\lambda\delta\leq C_\lambda\). All constants in this argument are obtained by finite rational arithmetic. The initial indices are covered by the definition of \(C_\lambda\).
\end{proof}

\begin{proposition}[The earlier \(3/2\)-scale length bound]\label{prop:23}
For \(\lambda=2\), so \(\theta=2/3\), the exact ruler interpretation above gives
\[
|R_M|<23M<258M
\]
for every dyadic \(M\).
\end{proposition}
\begin{proof}
The recurrence gives
\[
(a_0,\ldots,a_8)=(1,6,16,46,126,286,666,1586,3426),
\]
reproducing the finite values reported in the earlier preprint~\cite{xin}. In particular,
\[
(A_0,\ldots,A_7)=
\left(1,3,4,\frac{23}{4},\frac{63}{8},\frac{143}{16},
\frac{333}{32},\frac{793}{64}\right).
\]
For this explicit constant, split the recurrence at \(P=12\). The largest
child exponent in the finite initial contribution is
\[
H=\left\lfloor\frac{2(P-1)}3\right\rfloor=7,
\]
so the required initial values are \(A_0,\ldots,A_7\).
Grouping \(p=3t,3t+1,3t+2\) gives the exact tail
\[
\sum_{p\geq12}2^{\lfloor2p/3\rfloor-p}
=\sum_{t\geq4}\left(2^{-t}+2^{-t-1}+2^{-t-1}\right)
=\frac14.
\]
The exact finite initial contribution is
\[
3+\sum_{p=0}^{11}2^{\lfloor2p/3\rfloor-p}A_{\lfloor2p/3\rfloor}
=\frac{17647}{1024}.
\]
The eight initial values \(A_0,\ldots,A_7\) are all less than \(23\).
For \(h\geq8\), the terms with \(p<12\) are bounded by the displayed finite
initial contribution, and every child exponent in the remaining terms is
less than \(h\). Induction and the exact tail therefore give
\[
A_h\leq \frac{17647}{1024}+\frac{23}{4}
=\frac{23535}{1024}<23 .
\]
This completes the induction.
\end{proof}

\section{The historical auxiliary-series convergence}\label{app:series}
The earlier presentation used a positive tail series to isolate a finite initial contribution. Its convergence can be recorded without an implicit limiting argument.
For fixed \(\lambda\geq1\), define
\[
\begin{gathered}
(b_1,b_2,b_3)=(1,2^{2-\lambda/(\lambda+1)},4),\\
(\gamma_1,\gamma_2,\gamma_3)
=\left(\frac1\lambda,\frac{2\lambda+1}{\lambda(\lambda+1)},
\frac{\lambda+1}{\lambda}\right).
\end{gathered}
\]
For integers \(i\geq1\), the earlier tail is \(q_i^{\rm old}=\sum_{t=i}^{\infty}
\sum_{\nu=1}^3b_\nu2^{-\gamma_\nu t}\).
\begin{proposition}\label{prop:historical-series}
The sequence \(q_i^{\rm old}\) is strictly decreasing and tends to zero. In particular it has a first index with \(q_i^{\rm old}<1\).
\end{proposition}
\begin{proof}
All \(\gamma_\nu\) are positive. Each geometric sum converges, and
\[
q_i^{\rm old}=\sum_{\nu=1}^3
\frac{b_\nu2^{-\gamma_\nu i}}{1-2^{-\gamma_\nu}} .
\]
This expression tends to zero. Removing the positive \(t=i\) summand strictly decreases the tail. The set of positive indices with \(q_i^{\rm old}<1\) is nonempty, so well-ordering gives its first element.
\end{proof}
The main proof uses the direct summability condition on the new integer periods. Neither this historical auxiliary series nor the unguarded legacy words are needed for its arbitrary-window argument.

\section{A complete search procedure}\label{app:protocol}
Fix the square-log schedule or fixed rational parameters \(r,\eps\) before activation. Use a streaming cubic UES routine with the contract in Corollary~\ref{cor:low-space}: parameter \(N\) requests a word universal for every graph of size at most \(N\), every port numbering and every directed-edge start. The parameter is internal to the generator, not knowledge of the physical graph size.

Query the normalized skeleton by Lemma~\ref{lem:skeleton}. Its streaming routine needs neither a numerical exponent \(c\) nor the physical graph size. The skeleton symbols are the actual instructions supplied by that routine.

If the initial physical degree is zero, connectedness implies a single vertex; remain there and detect the treasure whenever it becomes active. Otherwise initialize \(p=0,a=1,t=1\). This is a valid fictitious virtual incoming state at \((v,0)\), whose physical tail need not have been visited.

At activation and at every subsequent slot boundary, check the current physical vertex for an active treasure before executing the next instruction. Halt if a meeting occurs. This includes a treasure already present at the agent's initial vertex.

At instruction index \(t\), choose the least dyadic \(M\geq t\). Query \(S_M[t]\) by Theorem~\ref{thm:table-free}, evaluating the boundary function and child lengths on demand in the procedure of Figure~\ref{fig:direct-query}. Apply the accumulated sign to the actual skeleton value modulo 3; formal indices are not physical instructions. The resulting value is \(b\).

At physical vertex \(v\) of degree \(\delta\), put \(z=(a+b)\bmod3\). Use the three updates from Proposition~\ref{prop:simulation}: for \(z=0\), stay and set \(p=(p-1)\bmod\delta,a=2\); for \(z=2\), stay and set \(p=(p+1)\bmod\delta,a=0\); for \(z=1\), traverse port \(p\), learn the destination entry port \(q\), and set \(p=q,a=1\). Each update consumes one slot. Increment \(t\) and repeat, checking for the treasure at the new slot boundary before the next instruction. The degree in a cycle update is the current vertex's degree; after traversal the stored physical index is the new entry port.

By induction, the physical position and \((p,a)\) equal the virtual directed-edge state after exactly \(t-1\) instructions. Prefix compatibility ensures that changing \(M\) for queries never restarts the infinite word. The logarithmic-workspace implementation retains no length table, expanded word or trajectory. The memoized implementation is an alternative when additional workspace is available.

If \(\tau\) instructions have been executed at the later activation, the next \(P\) instructions form positions \(\tau+1,\ldots,\tau+P\) of this same word. A cover-time choice of \(P\) for the \(2m\)-vertex replacement therefore visits every physical vertex while both agent and treasure are active. Stay slots are charged, so activation during virtual cycle motion introduces no uncounted delay.

Writing \(\Delta\) for the maximum physical degree only in the analysis, total workspace can be
\[
O\!\left(\log(t+2)+\log(\Delta+2)\right).
\]


\begin{thebibliography}{17}

\ifx \showCODEN    \undefined \def \showCODEN     #1{\unskip}     \fi
\ifx \showISBNx    \undefined \def \showISBNx     #1{\unskip}     \fi
\ifx \showISBNxiii \undefined \def \showISBNxiii  #1{\unskip}     \fi
\ifx \showISSN     \undefined \def \showISSN      #1{\unskip}     \fi
\ifx \showLCCN     \undefined \def \showLCCN      #1{\unskip}     \fi
\ifx \shownote     \undefined \def \shownote      #1{#1}          \fi
\ifx \showarticletitle \undefined \def \showarticletitle #1{#1}   \fi
\ifx \showURL      \undefined \def \showURL       {\relax}        \fi
\providecommand\bibfield[2]{#2}
\providecommand\bibinfo[2]{#2}
\providecommand\natexlab[1]{#1}
\providecommand\showeprint[2][]{arXiv:#2}
\makeatletter
\@ifundefined{NAT@parse@date}{}{\let\NAT@parse@date@orig\NAT@parse@date}
\@ifundefined{NAT@parse@date}{}{\def\NAT@parse@date#1#2#3#4#5#6@@{\NAT@parse@date@orig#1#2#3#4#5#6@@\def\NAT@tempyear{0000}\def\NAT@tempexlab{{?}}\ifx\NAT@year\NAT@tempyear\ifx\NAT@exlab\NAT@tempexlab\def\NAT@date{[n.\,d.]}\else\edef\NAT@date{[n.\,d.]\NAT@exlab}\fi\fi}}
\makeatother

\bibitem[Aleliunas et~al\mbox{.}(1979)]%
        {aleliunas}
\bibfield{author}{\bibinfo{person}{Romas Aleliunas},
  \bibinfo{person}{Richard~M. Karp}, \bibinfo{person}{Richard~J. Lipton},
  \bibinfo{person}{L{\'a}szl{\'o} Lov{\'a}sz}, {and} \bibinfo{person}{Charles
  Rackoff}.} \bibinfo{year}{1979}\natexlab{}.
\newblock \showarticletitle{Random Walks, Universal Traversal Sequences, and
  the Complexity of Maze Problems}. In \bibinfo{booktitle}{\emph{Proceedings of
  the 20th Annual Symposium on Foundations of Computer Science (FOCS)}}.
  \bibinfo{publisher}{IEEE}, \bibinfo{pages}{218--223}.
\newblock

\bibitem[Coppersmith et~al\mbox{.}(1993)]%
        {coppersmith}
\bibfield{author}{\bibinfo{person}{Don Coppersmith}, \bibinfo{person}{Prasad
  Tetali}, {and} \bibinfo{person}{Peter Winkler}.}
  \bibinfo{year}{1993}\natexlab{}.
\newblock \showarticletitle{Collisions among Random Walks on a Graph}.
\newblock \bibinfo{journal}{\emph{SIAM Journal on Discrete Mathematics}}
  \bibinfo{volume}{6}, \bibinfo{number}{3} (\bibinfo{year}{1993}),
  \bibinfo{pages}{363--374}.
\newblock

\bibitem[{De Marco} et~al\mbox{.}(2006)]%
        {demarco}
\bibfield{author}{\bibinfo{person}{Gianluca {De Marco}}, \bibinfo{person}{Luisa
  Gargano}, \bibinfo{person}{Evangelos Kranakis}, \bibinfo{person}{Danny
  Krizanc}, \bibinfo{person}{Andrzej Pelc}, {and} \bibinfo{person}{Ugo
  Vaccaro}.} \bibinfo{year}{2006}\natexlab{}.
\newblock \showarticletitle{Asynchronous Deterministic Rendezvous in Graphs}.
\newblock \bibinfo{journal}{\emph{Theoretical Computer Science}}
  \bibinfo{volume}{355}, \bibinfo{number}{3} (\bibinfo{year}{2006}),
  \bibinfo{pages}{315--326}.
\newblock

\bibitem[Dessmark et~al\mbox{.}(2006)]%
        {dessmark06}
\bibfield{author}{\bibinfo{person}{Anders Dessmark}, \bibinfo{person}{Pierre
  Fraigniaud}, \bibinfo{person}{Dariusz~R. Kowalski}, {and}
  \bibinfo{person}{Andrzej Pelc}.} \bibinfo{year}{2006}\natexlab{}.
\newblock \showarticletitle{Deterministic Rendezvous in Graphs}.
\newblock \bibinfo{journal}{\emph{Algorithmica}} \bibinfo{volume}{46},
  \bibinfo{number}{1} (\bibinfo{year}{2006}), \bibinfo{pages}{69--96}.
\newblock

\bibitem[Dessmark et~al\mbox{.}(2003)]%
        {dessmark03}
\bibfield{author}{\bibinfo{person}{Anders Dessmark}, \bibinfo{person}{Pierre
  Fraigniaud}, {and} \bibinfo{person}{Andrzej Pelc}.}
  \bibinfo{year}{2003}\natexlab{}.
\newblock \showarticletitle{Deterministic Rendezvous in Graphs}. In
  \bibinfo{booktitle}{\emph{Algorithms -- ESA 2003}}.
  \bibinfo{publisher}{Springer}, \bibinfo{pages}{184--195}.
\newblock

\bibitem[Disser et~al\mbox{.}(2019)]%
        {disser19}
\bibfield{author}{\bibinfo{person}{Yann Disser}, \bibinfo{person}{Jan
  Hackfeld}, {and} \bibinfo{person}{Max Klimm}.}
  \bibinfo{year}{2019}\natexlab{}.
\newblock \showarticletitle{Tight Bounds for Undirected Graph Exploration with
  Pebbles and Multiple Agents}.
\newblock \bibinfo{journal}{\emph{J. ACM}} \bibinfo{volume}{66},
  \bibinfo{number}{6}, Article \bibinfo{articleno}{40} (\bibinfo{date}{Oct.}
  \bibinfo{year}{2019}), \bibinfo{numpages}{41}~pages.
\newblock
\href{https://doi.org/10.1145/3356883}{doi:\nolinkurl{10.1145/3356883}}

\bibitem[Kouck{\'y}(2002)]%
        {koucky}
\bibfield{author}{\bibinfo{person}{Michal Kouck{\'y}}.}
  \bibinfo{year}{2002}\natexlab{}.
\newblock \showarticletitle{Universal Traversal Sequences with Backtracking}.
\newblock \bibinfo{journal}{\emph{J. Comput. System Sci.}}
  \bibinfo{volume}{65}, \bibinfo{number}{4} (\bibinfo{year}{2002}),
  \bibinfo{pages}{717--726}.
\newblock

\bibitem[Kowalski and Malinowski(2006)]%
        {kowalski06}
\bibfield{author}{\bibinfo{person}{Dariusz~R. Kowalski} {and}
  \bibinfo{person}{Adam Malinowski}.} \bibinfo{year}{2006}\natexlab{}.
\newblock \showarticletitle{How to Meet in Anonymous Network}. In
  \bibinfo{booktitle}{\emph{Structural Information and Communication Complexity
  (SIROCCO 2006)}} \emph{(\bibinfo{series}{Lecture Notes in Computer Science},
  Vol.~\bibinfo{volume}{4056})}. \bibinfo{publisher}{Springer},
  \bibinfo{pages}{44--58}.
\newblock

\bibitem[Kowalski and Pelc(2004)]%
        {kowalski04}
\bibfield{author}{\bibinfo{person}{Dariusz~R. Kowalski} {and}
  \bibinfo{person}{Andrzej Pelc}.} \bibinfo{year}{2004}\natexlab{}.
\newblock \showarticletitle{Polynomial Deterministic Rendezvous in Arbitrary
  Graphs}. In \bibinfo{booktitle}{\emph{Algorithms and Computation (ISAAC
  2004)}} \emph{(\bibinfo{series}{Lecture Notes in Computer Science},
  Vol.~\bibinfo{volume}{3341})}. \bibinfo{publisher}{Springer},
  \bibinfo{pages}{644--656}.
\newblock

\bibitem[Pelc(2023)]%
        {pelc23}
\bibfield{author}{\bibinfo{person}{Andrzej Pelc}.}
  \bibinfo{year}{2023}\natexlab{}.
\newblock \bibinfo{title}{Deterministic Rendezvous Algorithms}.
\newblock \bibinfo{howpublished}{arXiv:2303.10391}.
\newblock
\urldef\tempurl%
\url{https://arxiv.org/abs/2303.10391}
\showURL{%
\tempurl}
\newblock
\shownote{Survey preprint}.

\bibitem[Reingold(2005)]%
        {reingold05}
\bibfield{author}{\bibinfo{person}{Omer Reingold}.}
  \bibinfo{year}{2005}\natexlab{}.
\newblock \showarticletitle{Undirected {ST}-Connectivity in Logspace}. In
  \bibinfo{booktitle}{\emph{Proceedings of the Thirty-Seventh Annual ACM
  Symposium on Theory of Computing (STOC)}}. \bibinfo{publisher}{ACM},
  \bibinfo{pages}{376--385}.
\newblock

\bibitem[Reingold(2008)]%
        {reingold08}
\bibfield{author}{\bibinfo{person}{Omer Reingold}.}
  \bibinfo{year}{2008}\natexlab{}.
\newblock \showarticletitle{Undirected Connectivity in Log-Space}.
\newblock \bibinfo{journal}{\emph{J. ACM}} \bibinfo{volume}{55},
  \bibinfo{number}{4}, Article \bibinfo{articleno}{17} (\bibinfo{year}{2008}),
  \bibinfo{numpages}{24}~pages.
\newblock
\href{https://doi.org/10.1145/1391289.1391291}{doi:\nolinkurl{10.1145/1391289.1391291}}

\bibitem[Sudo et~al\mbox{.}(2026)]%
        {sudo}
\bibfield{author}{\bibinfo{person}{Yuichi Sudo}, \bibinfo{person}{Fukuhito
  Ooshita}, {and} \bibinfo{person}{Sayaka Kamei}.}
  \bibinfo{year}{2026}\natexlab{}.
\newblock \showarticletitle{Self-Stabilizing Graph Exploration by a Single
  Agent}.
\newblock \bibinfo{journal}{\emph{Theoretical Computer Science}}
  \bibinfo{volume}{1082} (\bibinfo{date}{Sept.} \bibinfo{year}{2026}),
  \bibinfo{pages}{116085}.
\newblock
\href{https://doi.org/10.1016/j.tcs.2026.116085}{doi:\nolinkurl{10.1016/j.tcs.2026.116085}}

\bibitem[Ta-Shma and Zwick(2007)]%
        {tz07}
\bibfield{author}{\bibinfo{person}{Amnon Ta-Shma} {and} \bibinfo{person}{Uri
  Zwick}.} \bibinfo{year}{2007}\natexlab{}.
\newblock \showarticletitle{Deterministic Rendezvous, Treasure Hunts and
  Strongly Universal Exploration Sequences}. In
  \bibinfo{booktitle}{\emph{Proceedings of the Eighteenth Annual ACM-SIAM
  Symposium on Discrete Algorithms (SODA)}}. \bibinfo{publisher}{Society for
  Industrial and Applied Mathematics}, \bibinfo{pages}{599--608}.
\newblock

\bibitem[Ta-Shma and Zwick(2014)]%
        {tz}
\bibfield{author}{\bibinfo{person}{Amnon Ta-Shma} {and} \bibinfo{person}{Uri
  Zwick}.} \bibinfo{year}{2014}\natexlab{}.
\newblock \showarticletitle{Deterministic Rendezvous, Treasure Hunts, and
  Strongly Universal Exploration Sequences}.
\newblock \bibinfo{journal}{\emph{ACM Transactions on Algorithms}}
  \bibinfo{volume}{10}, \bibinfo{number}{3}, Article \bibinfo{articleno}{12}
  (\bibinfo{year}{2014}), \bibinfo{numpages}{15}~pages.
\newblock
\href{https://doi.org/10.1145/2601068}{doi:\nolinkurl{10.1145/2601068}}

\bibitem[Xin(2007)]%
        {xin07}
\bibfield{author}{\bibinfo{person}{Qin Xin}.} \bibinfo{year}{2007}\natexlab{}.
\newblock \showarticletitle{Faster Treasure Hunt and Better Strongly Universal
  Exploration Sequences}. In \bibinfo{booktitle}{\emph{Algorithms and
  Computation (ISAAC 2007)}} \emph{(\bibinfo{series}{Lecture Notes in Computer
  Science}, Vol.~\bibinfo{volume}{4835})},
  \bibfield{editor}{\bibinfo{person}{Takeshi Tokuyama}} (Ed.).
  \bibinfo{publisher}{Springer}, \bibinfo{pages}{549--560}.
\newblock
\href{https://doi.org/10.1007/978-3-540-77120-3_48}{doi:\nolinkurl{10.1007/978-3-540-77120-3_48}}

\bibitem[Xin(2012)]%
        {xin}
\bibfield{author}{\bibinfo{person}{Qin Xin}.} \bibinfo{year}{2012}\natexlab{}.
\newblock \bibinfo{title}{Faster Treasure Hunt and Better Strongly Universal
  Exploration Sequences}.
\newblock \bibinfo{howpublished}{arXiv:1204.5442v1}.
\newblock
\urldef\tempurl%
\url{https://arxiv.org/abs/1204.5442}
\showURL{%
\tempurl}
\newblock
\shownote{Preprint}.

\end{thebibliography}
\end{document}